\documentclass [11pt,twoside,a4paper]{article}
\usepackage{shrthnds}
\usepackage{cite}
\usepackage{graphicx}
\usepackage[usenames]{color}
\usepackage{subfigure}
\usepackage{setspace}
\usepackage[hang,small]{caption}
\usepackage{amsmath,amssymb}
\usepackage{slashed}                  
\usepackage{geometry}
\usepackage{fancyhdr}
\usepackage{titlesec,titletoc}
  \titleformat{\section}{\Large\sf\bfseries}{\thesection}{1em}{}
  \titleformat{\subsection}{\large\sf\bfseries}{\thesubsection}{1em}{}
\newcommand{\auth}[2]{{\large #1}\footnote{email:~#2}}
\newcommand{\affS}[1]{{\it #1}}
\newcommand{\verttle}[2]{\vspace{-1cm}\begin{flushright}{\small #1}\end{flushright}\vspace{0.5cm} {\sf\bfseries #2}}
\newcommand{\nabstract}[2][]{\bc\begin{minipage}{0.9\textwidth}\begin{spacing}{1}{\small {\sf\bfseries Abstract:} #2 }\end{spacing}
#1 \end{minipage}\ec}
\newcommand{\nkeywords}[1][]{~\\\small{ {\sf\bfseries Keywords:} #1 }}

\usepackage{ulem}

\title{\verttle{LPT-ORSAY-12-112}{Effect of Anomalous Couplings on the Associated Production of a Single Top Quark and a Higgs Boson at the LHC\\}}
\author{
  \auth{Pankaj Agrawal $^a$}{agrawal@iopb.res.in}~,
  \auth{Subhadip Mitra $^b$}{subhadip.mitra@th.u-psud.fr}~~and
  \auth{Ambresh Shivaji $^c$}{ambreshkshivaji@hri.res.in}\\~\\  
  \affS{a) Institute of Physics, Saink School Post, Bhubaneswar 751 005, India}\\
  \affS{b) Laboratoire de Physique Th\'{e}orique, CNRS - UMR 8627,}\\
  \affS{B\^{a}timent 210, Univ. Paris-Sud 11, F-91405 Orsay Cedex, France}\\
  \affS{c) Regional Centre for Accelerator-based Particle Physics,} \\
  \affS{Harish-Chandra Research Institute, Chhatnag Road, Jhusi, Allahabad 211019, India}\\~\\
}
\newcommand{\pghdr}{\footnotesize P. Agrawal {\it et al.} -- Effect of Anomalous Couplings on the Associated Production \ldots}
\date{\today}

\begin{document}

\maketitle

\nabstract[\nkeywords{Single top, Higgs boson, LHC, Anomalous couplings}]{
We consider the production of a single top quark in association with a Higgs 
boson at the LHC. In particular, we compute the cross sections for the processes  
$ p p  \to  t h j$, $t h b$, $t h W$, $t h j j$, $t h j b$, $t h W j$, $t h W b$ 
in the presence of the anomalous $Wtb, WWh$ and $tth$ couplings. We find that 
the anomalous $Wtb$ and $tth$ couplings can enhance the cross sections significantly. 
We also analyze a few signatures and show that, if these couplings are indeed anomalous, 
then with enough data, one should be able to observe the production of the 
Higgs boson in association with single top quark. }
\bigskip

\section{Introduction}\label{sec:introduction}

    So far the Standard Model (SM) has been remarkably successful in explaining the 
    data from the modern hadron colliders like the Tevatron at Fermilab or the Large Hadron
    Collider (LHC) at CERN. We have now very strong indications that the only missing piece 
    of the SM, the Higgs boson, has been discovered \cite{HiggsAtlas,HiggsCMS}. On the other hand, 
   there does not seem to be any stand-out signal of any of the beyond the 
    Standard Model (BSM) scenarios. There exist wide variety of scenarios with
   specific signatures to validate them. Some of these scenarios have
   overlapping signatures. Therefore, even if one finds a new signal, it may
   require a lot of work to ensure the connection of the signal with a
   specific model. This suggests that, apart from the model-specific analysis of the data, it will 
   also be useful to look for BSM scenarios in model independent ways. One
   method to do so is by constructing suitable effective Lagrangians. These effective 
   Lagrangians have terms that are consistent 
   with some of the aspects of the SM, in particular symmetries, but contain
   higher dimensional (non-renormalizable) operators. Because of the non-renormalizable
   nature of the extra terms, these effective Lagrangians can only be
   used in a restricted domain of the energy scale. The particle content of
   these effective Lagrangian models is same as that of the SM.
   The extra terms in the Lagrangian can introduce new interactions, or they 
   can modify the existing interactions of some of the particles. In particular, 
   we note that, we can have modifications of the $Wtb, tth$ and $WWh$ interactions
   that can be parametrized as anomalous couplings.

   After the discovery of the Higgs boson at the LHC, it would be 
   important to study various properties of it. In particular,
   one would like to study the production of the Higgs boson via all possible
   channels. One such category of channels is the production of a Higgs
   boson in association with single top quark. In these processes, there
   can be additional particles, apart from a top quark and a Higgs boson.
   Some of these processes have been studied within the context of the SM \cite{Maltoni},
   and also considering scaled up $tth$ and $WWh$ couplings \cite{Barger:2009ky}.
   These processes are similar to the single top-quark production processes. 
   In this case, a Higgs boson is emitted either from the top quark or the $W$ boson. 
   Due to the similarity with the single top-quark production processes, one would 
   expect these processes to contribute significantly to the Higgs boson production 
   at the LHC. However, as pointed out in Ref. \cite{Maltoni}, for the Higgs boson 
   mass, $m_h <$ 200 GeV, the cross sections of such processes turn out to be rather 
   small compared to what is expected from the single top-quark production at the LHC.  
   At the LHC, for $m_h \sim 100-150$ GeV, the dominant contributions come from the 
   $t$-channel $W$ exchange process, $ p p \to thj$ and associated production with 
   a $W$ boson, $p p \to tWh$. The authors of Ref. \cite{Maltoni} demonstrated that 
   for both of these channels, there is a destructive interference between the 
   diagrams where the Higgs boson is emitted from the top quark and ones with 
   the Higgs boson emitted from the $W$ boson. Because of the small cross sections, 
   these channels are generally not considered as significant to measure the 
   properties of the Higgs boson. However, inclusion of the anomalous couplings changes 
   the picture. The cross sections can be significantly enhanced to make these 
   processes phenomenologically useful. In this paper, we study the effect of 
   anomalous $Wtb, tth$ and $WWh$ interactions on the cross sections and distributions 
   of the processes that involve the production of  a single top quark in association 
   with a Higgs boson at the LHC. We find that the enhancement in the cross sections 
   can be more than a factor of ten for some values of the $Wtb$ and $tth$ anomalous 
   couplings, and as a result 
   the associated production of a single top quark with the Higgs boson
   can become significant at the LHC.  Since the associated production of a 
   Higgs boson with a top quark is quite suppressed in the SM and, at the same 
   time, very sensitive to some anomalous couplings, it can provide us a 
   new opportunity to probe any new physics model that can generate these 
   anomalous couplings. Therefore, once observed, these channels can not only
   give us useful information about the couplings but also help us to identify 
   or constrain some new physics models. However, in this paper we shall not 
   pursue the details of the possible new physics models, rather restrict 
   ourselves to the study of the effect of the anomalous couplings that can 
   appear in the $Wtb, tth$ and $WWh$ vertices on the $pp\to thX$ process in 
   the effective theory framework. Recently, there have also been a few studies
   that consider the change in the sign of the $tth$ Yukawa coupling on the
   associated production of a single top quark and a Higgs boson
   \cite{Biswas:2012bd,Farina:2012xp,Biswas:2013bd}. This change of sign
   leads to a constructive interference among the diagrams and thus a
   significant increase in the $thj$ and $thbj$ cross sections. It is argued
   that this enhancement can be detected at the LHC using various
   decay modes of the Higgs boson \cite{Biswas:2012bd,Farina:2012xp,Biswas:2013bd}. 
   In this paper, we are not only
   considering this situation, but general anomalous $tth$ coupling. In
   addition, we consider the effect of anomalous $tbW$ and $WWh$ couplings
   also. We also consider a few signatures of the single top
   quark and a Higgs boson production and show that these signatures could
   be visible at the LHC. 

   The organization of the paper is as follows. In section 2, we describe the 
   processes under consideration. In section 3, we discuss the anomalous $Wtb, tth$ 
   and $WWh$ couplings. In section 4, we present the numerical results. In section 5, 
   we discuss the possibility of observing these processes at the LHC. In the last 
   section, we present our conclusions.

\section{Processes}

 \begin{figure}[!h]
\bc
\subfigure[]{\includegraphics [angle=0,width=.15\linewidth] {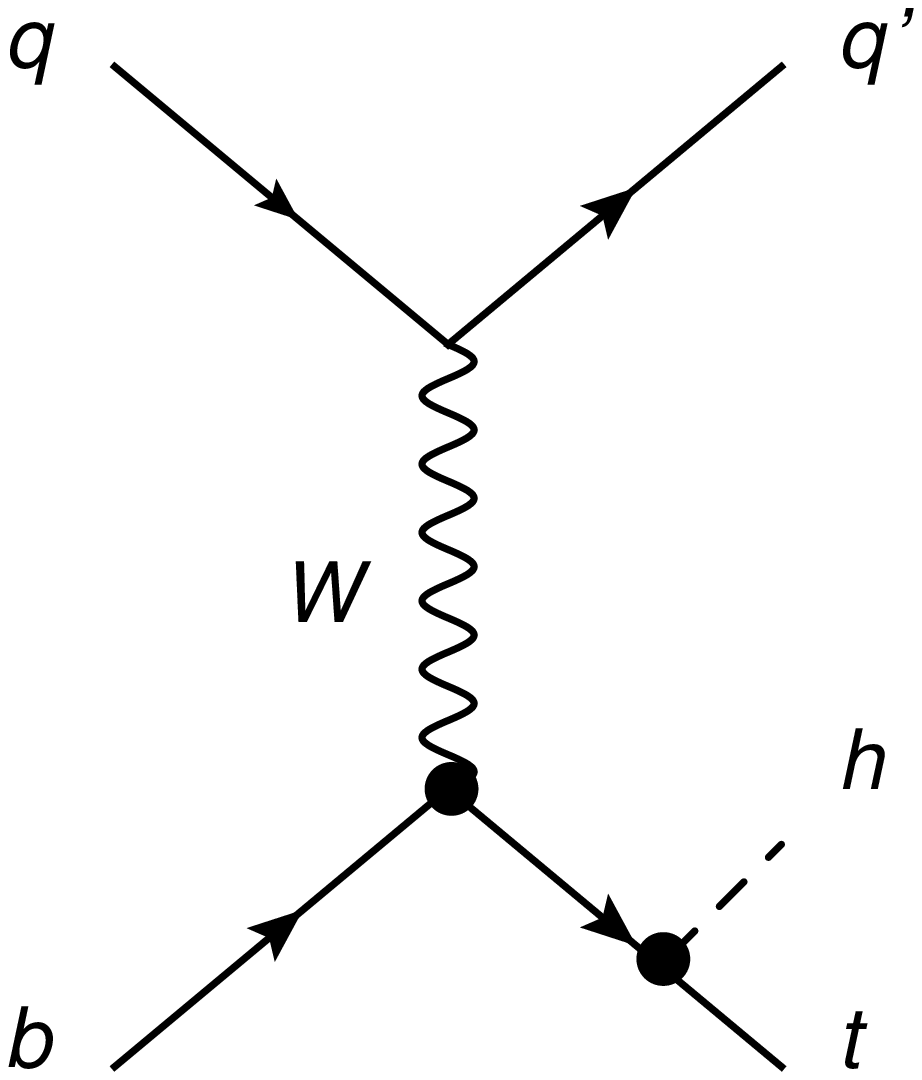}}
\subfigure[]{\includegraphics [angle=0,width=.24\linewidth]{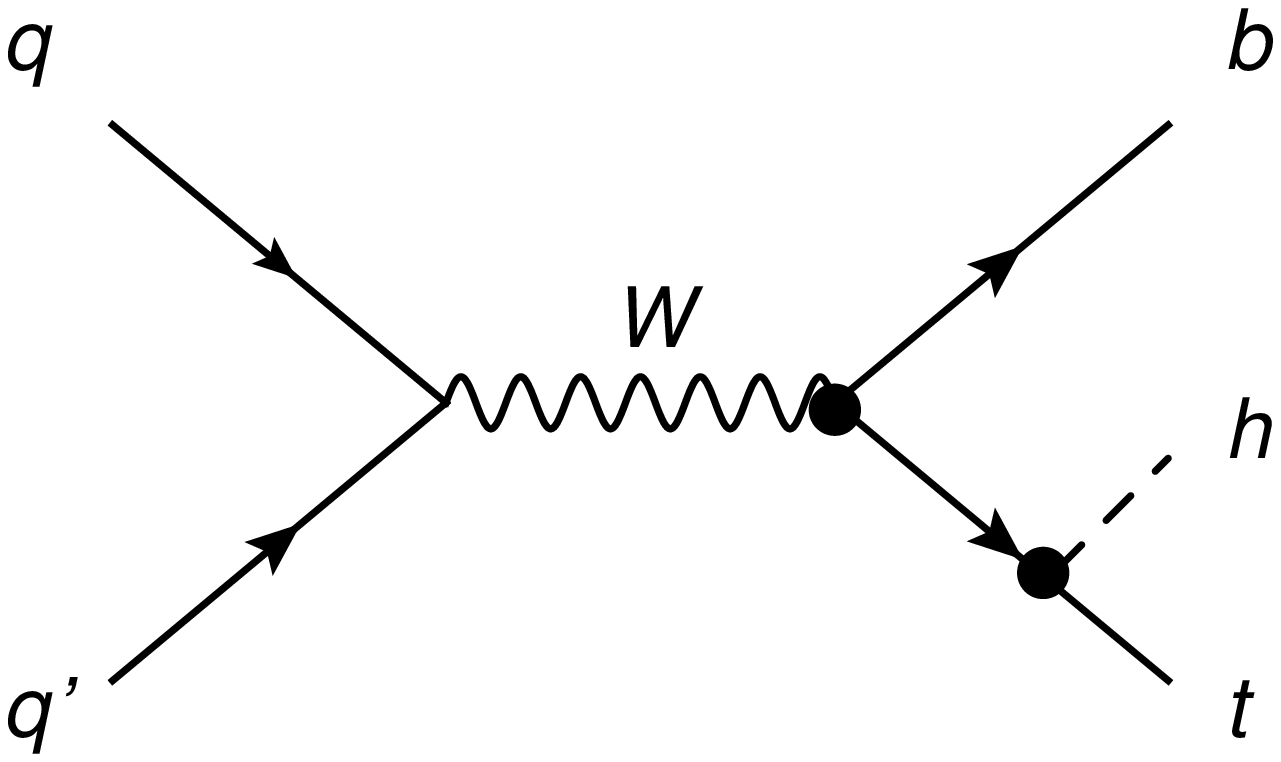}}
\subfigure[]{\includegraphics [angle=0,width=.24\linewidth]{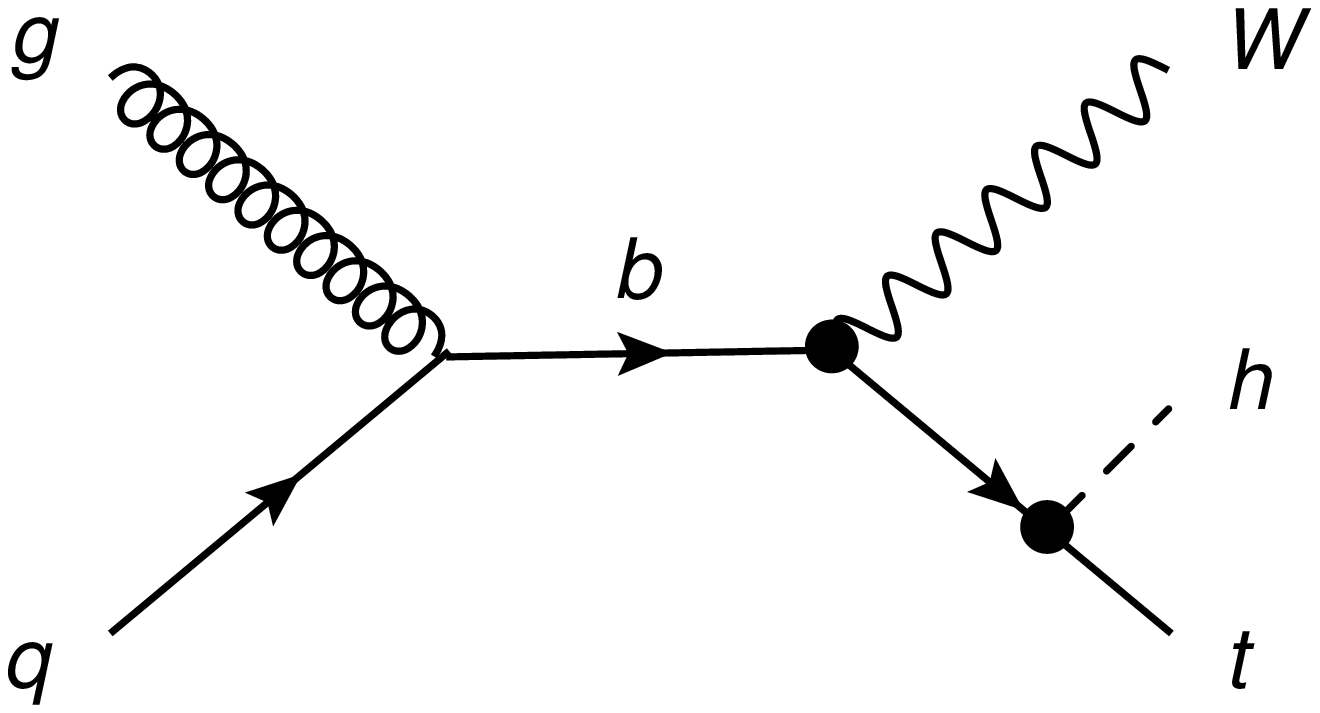}}\\
\subfigure[]{\includegraphics [angle=0,width=.15\linewidth]{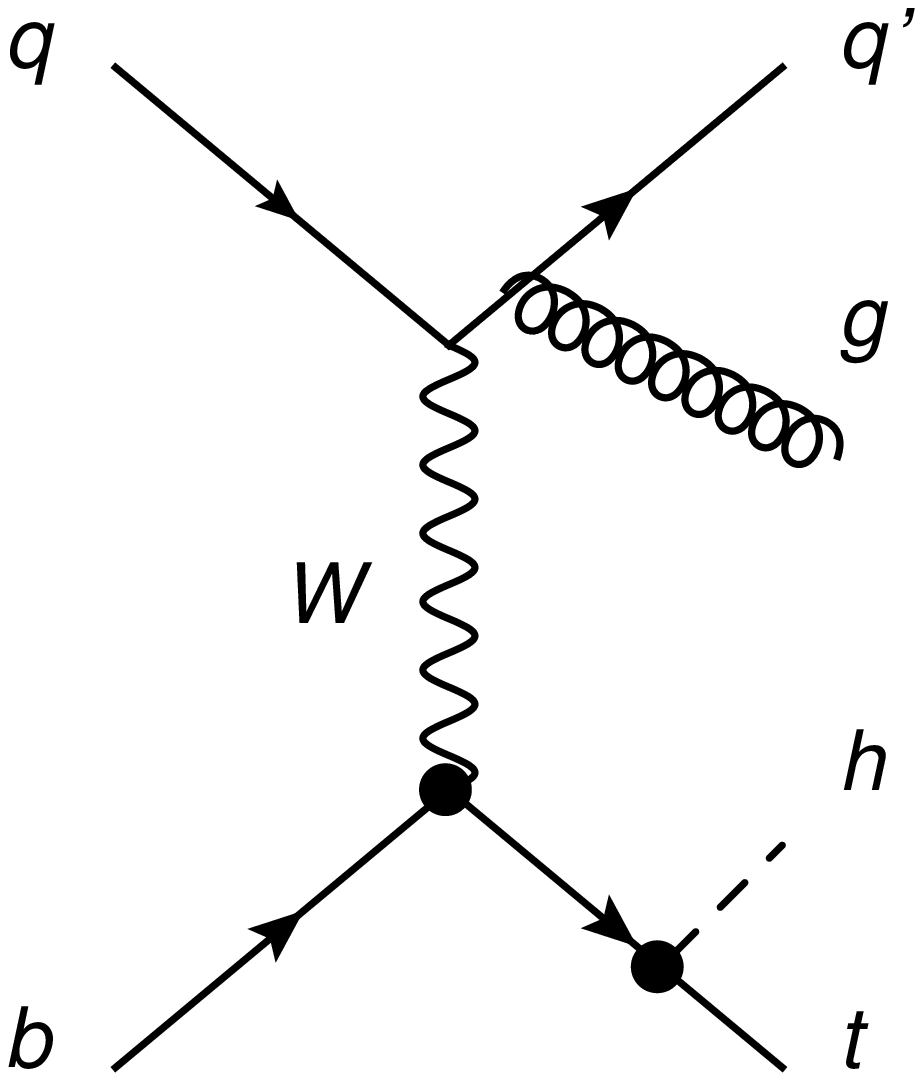}}\\
\subfigure[]{\includegraphics [angle=0,width=.24\linewidth]{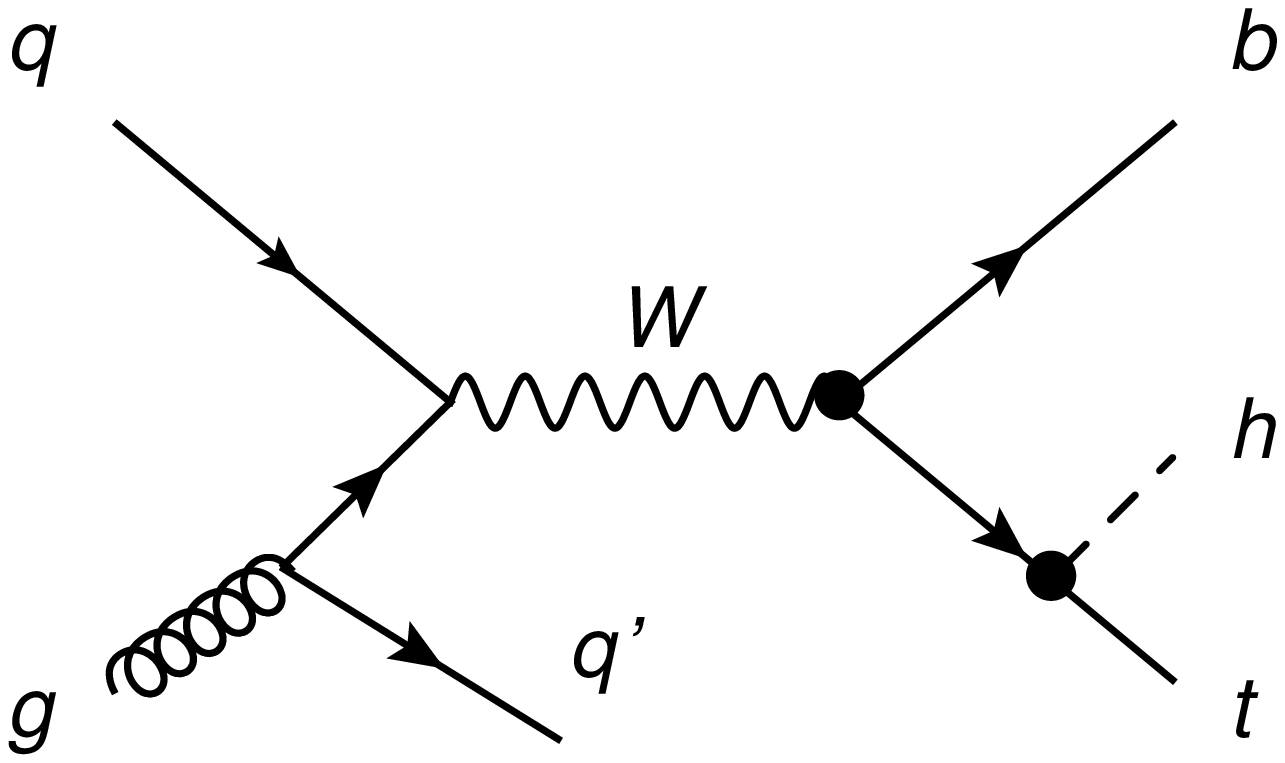}}
\subfigure[]{\includegraphics [angle=0,width=.24\linewidth]{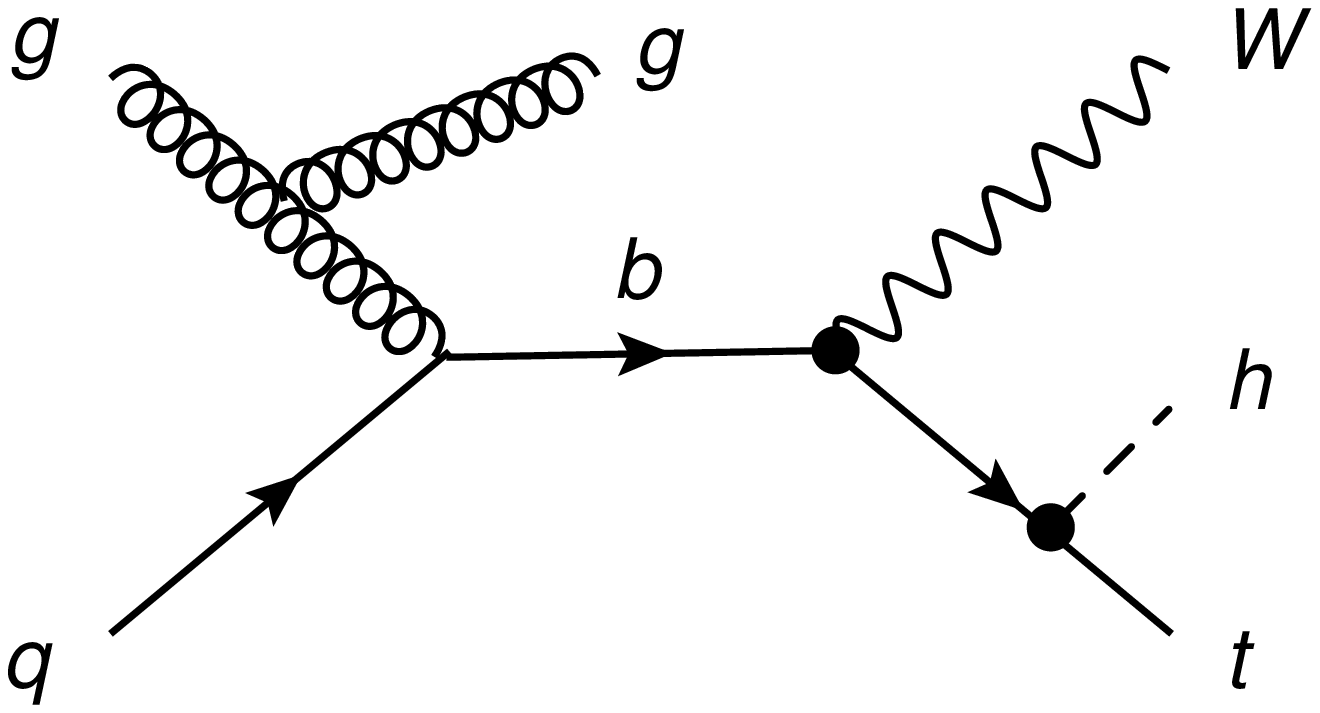}}
\subfigure[]{\includegraphics [angle=0,width=.24\linewidth]{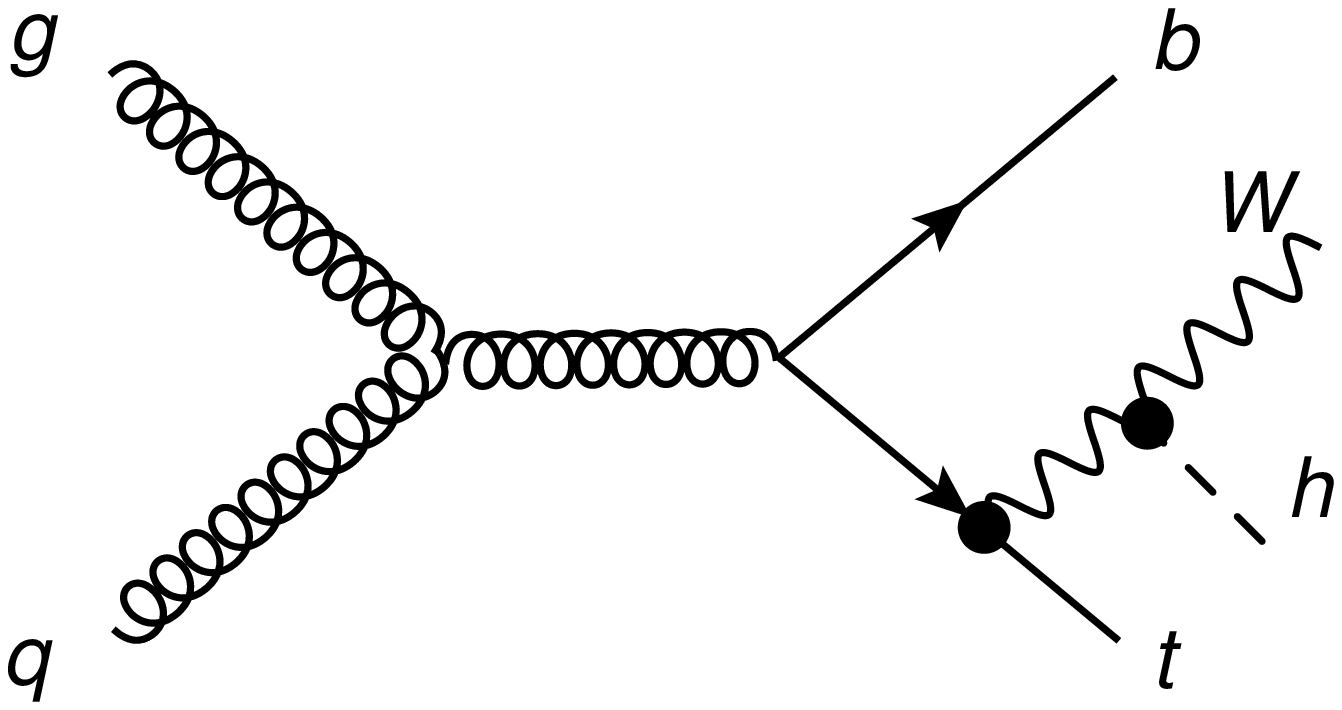}}
\ec
\caption{Representative Feynman diagrams for the processes listed in Eqs. \ref{eq:thj} - \ref{eq:thwb}.
\label{fig:total}}
\end{figure}

In this section, we describe those processes for the production of a Higgs boson where it is
   produced in association with a single top quark. In our analysis we include the tree-level 
   leading order and the subleading order processes ({\it i.e.,} processes with an extra jet) 
   that have significant cross sections. The leading order processes are following
\begin{eqnarray}
 p ~p & \to &  t ~h ~j ~X, \label{eq:thj}\\
 p ~p & \to &  t ~h ~b ~X, \label{eq:thb}\\
 p ~p & \to &  t ~h ~W ~X \label{eq:thw}\end{eqnarray}
 and the processes with an extra jet are, 
 \ba
 p ~p & \to &  t ~h ~j ~j ~X, \label{eq:thjj}\\
 p ~p & \to &  t ~h ~j ~b ~X, \label{eq:thjb}\\
 p ~p & \to &  t ~h ~W ~j ~X, \label{eq:thwj}\\
 p ~p & \to &  t ~h ~W ~b ~X. \label{eq:thwb}
\ea
  Here `$j$' represents a jet from a light quark (excluding bottom quark) or a gluon. 
  Representative parton level diagrams are displayed in Fig. \ref{fig:total}. 
  The leading order processes can be classified into three categories: 
   \begin{enumerate}
    \item process with $W$ boson in $t$-channel, $p p \to  t h j$,
    \item process with $W$ boson in $s$-channel, $p p \to  t h b$ and    
    \item process with $W$ boson in the final state, $p p \to  t h W$.
   \end{enumerate}
   As we shall see, the $t$-channel process has the largest cross section, while the $s$-channel 
   process has the smallest cross section. The subleading diagrams can be obtained by 
   adding an extra jet (either light or $b$-jet) to these three processes. Some of these
   subleading processes can have cross sections larger than the leading-order
   processes, specially the s-channel leading-order process. All the above processes 
   contain one $tbW$ vertex and one $tth$ or $WWh$ vertex. That is why we study the 
   effect of anomalous couplings in these vertices on the cross sections.

  Although subleading processes can  have relatively significant 
  cross sections, but one has to be careful while computing their contribution 
  at the matrix-element level. These extra jets can be soft and thus lead to infrared divergences.
  To avoid the soft jet contribution one has to set a reasonably large $p_T$ cut for them.
  Apart from this, there is also the possibility of over counting. Like, {\it e.g.},
  in the case of the process $ p p \to   t h j j $, the jet pair can come from an on-shell $W$ decay 
  making it a $ p p \to t h W $ process. Hence to estimate the cross section of this process we 
  don't allow any on-shell $W$. Similarly, for the process $ p p \to t h W b $, the $bW$ pair 
  can come from the decay of an on-shell top quark. However, in that case the actual process 
  will be $pp \to tth$, which has a much larger cross section than the $th$ production. 
  To avoid such a situation, in our calculation, we allow only one of the top quark to go on-shell.

\section{Anomalous Interactions}

   As we discussed above, the processes under consideration have three
   electroweak vertices - $tbW, tth$, and $WWh$. (Since $Wqq^\prime$ vertex 
   with $q$ and $q^\prime$ being the light quarks is severely constrained,
   we don't include the possibility of this vertex being anomalous.)
   We consider the general modification of these vertices due to BSM interactions.
   The possible general structure of these vertices have been extensively
   discussed in the literature \cite{AguilarSaavedra:2008zc,Saavedra1,Hagiwara:1993ck,
   Barger:2003rs,Whisnant:1994fh}. One parametrizes the effect of heavy BSM physics 
   by introducing the most general independent set of higher dimensional operators 
   that satisfies the gauge symmetries of the SM. However, some of these terms generally 
   reduce to simpler and more familiar forms when relations such as the equations 
   of motion of the fields are used. We will use these simpler forms for our 
   calculations.

\subsection*{Anomalous Couplings in the $tbW$ Vertex}
   In the SM, the $tbW$ coupling is {\em V-A} type. Therefore, only the left-handed 
   fermion fields couple to the $W$ boson. So, it allows only a 
   left-handed top quark to decay into a bottom quark and a $W$ boson. However, 
   BSM physics can generate several other possible $tbW$ couplings.
   One can write down the most general $tbW$ interaction 
   that includes corrections from dimension-six operators \cite{AguilarSaavedra:2008zc},
\ba
\mc L_{tbW} = \frac{g}{\sqrt2}\bar b\left[\g^\m\left(f_{1L}P_{L} + f_{1R}P_{R}\right) W^-_\m  
+\frac{\s^{\m\n}}{m_W}(f_{2L}P_{L} + f_{2R}P_{R}) \left(\pr_{\n}W^-_\m\right)\right]t + H.c.,\label{eq:tbw_lag}
\ea
where, in general, $f_{iL/R}$'s are complex dimensionless parameters. Also 
$P_{L,R} = {1 \over 2} (1 \mp \gamma_{5})$. In the SM, $f_{1L} = V_{tb} \approx 1$ 
while $f_{1R} = f_{2L} = f_{2R} = 0$.  In our analysis, we assume the $f_{iL/R}$'s 
to be real for simplicity.

Both recent LHC data and Tevatron data put bound on these parameters. Till now 
Tevatron puts more stringent bound on these as compared to the LHC\cite{Saavedra:LHC_Tevatron}. 
The Tevatron bounds are roughly
\ba
0.8  \lesssim &f_{1L} &  \lesssim 1.2\,,\nn\\ 
-0.5  \lesssim &f_{1R}&  \lesssim 0.5\,,\\
-0.2 \lesssim &f_{2L/R} &  \lesssim 0.2\,.\nn
\ea
Notice that these bounds are quite loose. Therefore, the SM results can have significant corrections. 
We note that there are also bounds on these parameters from the top-quark decays \cite{Drobnak:2010ej},
which are not more stringent.

\subsection*{Anomalous Couplings in the $tth$ Vertex}
In the SM, the top quark couples with the Higgs boson via the Yukawa coupling. In the effective theory, 
the most general vertex for $tth$ interaction can be parametrized as \cite{Saavedra1},
\ba
\mc L_{\bar tth} = -\frac{m_t}{v}\bar t\left[\left(1+ y_t^V\right)  + i y_t^A \g_5\right]t h.\label{eq:tth_lag}
\ea
In the SM, $y_t^V = y_t^A = 0$ and the first non-zero contributions to $y_t^V$ and  $y_t^A$ come from 
dimension six operators.

So far there is no direct experimental measurement of the top-quark Yukawa couplings. 
However, from the production of the Higgs boson at the LHC through the $ g g \to h$
process, one can obtain information about the $tth$ vertex. The recent analyses of 
the Higgs boson production and decays generally assume a generic scaling behavior of 
the top-quark Yukawa coupling (see, {\it e.g.,} \cite{Falkowski:2013dza}),
\ba
\mc L_{\bar tth} = -C_t\frac{m_t}{v}\bar tt h.
\ea
The coupling $C_t$ can be written in our notation as,
\ba
C_t = y_t^V + 1.
\ea
These analyses indicate that the value of $C_t$ is close to 1.
However, the uncertainty in these estimates 
still leaves some freedom for the anomalous coupling in the $tth$ vertex.
From the theoretical side, unitarity constraints allow order one values 
for $y_t^V$ and  $y_t^A$ \cite{Whisnant:1994fh}. We note that there has been
a recent bound on these Yukawa couplings by considering the production of a 
Higgs boson \cite{Nishiwaki:2013cma}. To estimate the observability, we have restricted our analysis by the bounds of this study.

\subsection*{Anomalous Couplings in the $WWh$ Vertex}

The new higher dimensional operators that can contribute to $WWh$ Vertex can be 
written as\cite{Hagiwara:1993ck,Barger:2003rs}
\ba
\mc L_{WWh} &=& g^1_{Wh} \left(G_{\m\n}^+W^{-\m} + G_{\m\n}^-W^{+\m} \right)\pr^\n h + g^2_{Wh}\left(G_{\m\n}^-G^{+\m\n}\right) h \nn\\
&&- g^3_{Wh}\frac{m_W^2}{v}\left(W_\m^+W^{-\m}\right)h,\label{eq:WWh_lag}
\ea
where
\ba
G_{\m\n}^\pm = \pr_\m W_\n^\pm - \pr_\n W_\m^\pm \pm i g \left(W^3_\m W^\pm_\n-W^3_\n W^\pm_\m\right). 
\ea
The third term in Eq. \ref{eq:WWh_lag} comes form the normalization of the Higgs boson kinetic 
term which gets modified due to higher dimensional operators. The constraints coming from 
the electroweak precision data are\cite{Zhang:2003it},
\ba 
-0.16{\rm ~TeV}^{-1} \lesssim & g^1_{Wh} & \lesssim 0.13{\rm ~TeV}^{-1}\label{eq:lmtwwh1}\,,\\
-0.26{\rm ~TeV}^{-1} \lesssim & g^2_{Wh} & \lesssim 0.29{\rm ~TeV}^{-1}\,.\label{eq:lmtwwh2}
\ea
Like the $tth$ couplings, the present Higgs boson data from the LHC favors the 
SM values for the $WWh$ couplings. In Ref. \cite{Falkowski:2013dza} the authors 
indicate that the couplings of the Higgs boson to the $W$  boson lie within 20 \% of 
those of the SM values.

\section{Results}
\begin{figure}[!t]
\bc
\includegraphics [angle=-90,width=0.50\linewidth]{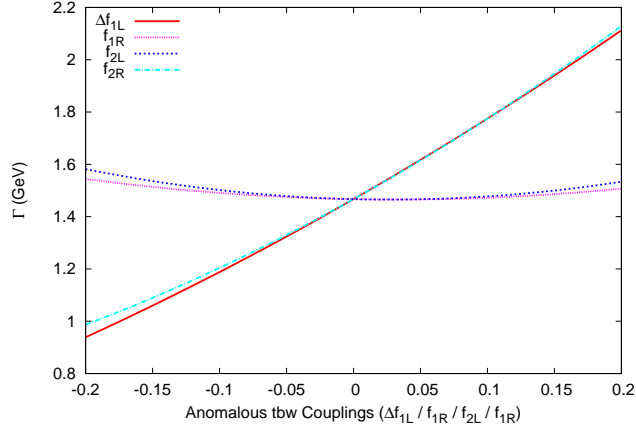}
\ec
\caption{Dependence of top quark width on the anomalous couplings present in the $tbW$ vertex (defined in Eq. \ref{eq:tbw_lag}) -- 
$\Dl f_{1L}= f_{1L} - 1$, $f_{1R}$, $f_{2L}$ and $f_{2R}$.}\label{fig:top_width}
\end{figure}
The main decay mode of the top quark is $t\to bW$ with a branching ratio of almost 99\%.
Therefore, the  presence of anomalous couplings in the $tbW$ vertex can modify the top quark
 width significantly. With anomalous couplings, the top quark width is

\begin{eqnarray}
 \Gamma(t \to b\;W) = \frac{G_F}{8\pi\sqrt{2}}\; m_t^3\;(1-x^2) 
                      \Big[ (1+x^2-2x^4)(f_{1L}^2+f_{1R}^2) \nonumber &\\
                           + (2-x^2-x^4)(f_{2L}^2+f_{2R}^2) 
                           + 6x(1-x^2)(f_{1L}f_{2R}+f_{2L}f_{1R})\Big],
\end{eqnarray}
where $x= M_W/m_t$.

In Fig. \ref{fig:top_width}, we show the dependence of the decay width of the top quark 
on $\Dl f_{1L}= f_{1L} - 1$, $f_{1R}$, $f_{2L}$ and $f_{2R}$. We see that the top quark width 
can change by about $\pm 50 \%$ on varying the values of $f_{1L}$ or $f_{2R}$. However, the width
is relatively immune to the change in the values of $f_{2L}$ or $f_{1R}$. We can understand this
as follows.  Since, $f_{1L} = 1 + \Delta$ and other couplings are $\sim \Delta$, this implies
\begin{equation}
f_{1L}^2 \simeq 1 + 2 \Delta; \; f_{1R}^2 = f_{2L}^2 = f_{2R}^2 = f_{2L}f_{1R} \simeq  \Delta^2 ; \;
 {\rm and} \; f_{1L}f_{2R} \simeq \Delta.  
\end{equation}
This explains the strong dependence of the decay width on $f_{1L}$ and $f_{2R}$. The weak 
dependence of the width on the couplings $f_{2L}$ and $f_{1R}$ is essentially
due to the absence of the terms proportional to $f_{1L}f_{1R}$ and $f_{1L}f_{2L}$.
One needs to include the modified widths when considering the decays of the top quark.

\begin{figure}[!h]
\bc
\subfigure[]{\includegraphics [angle=-90,width=.49\linewidth] {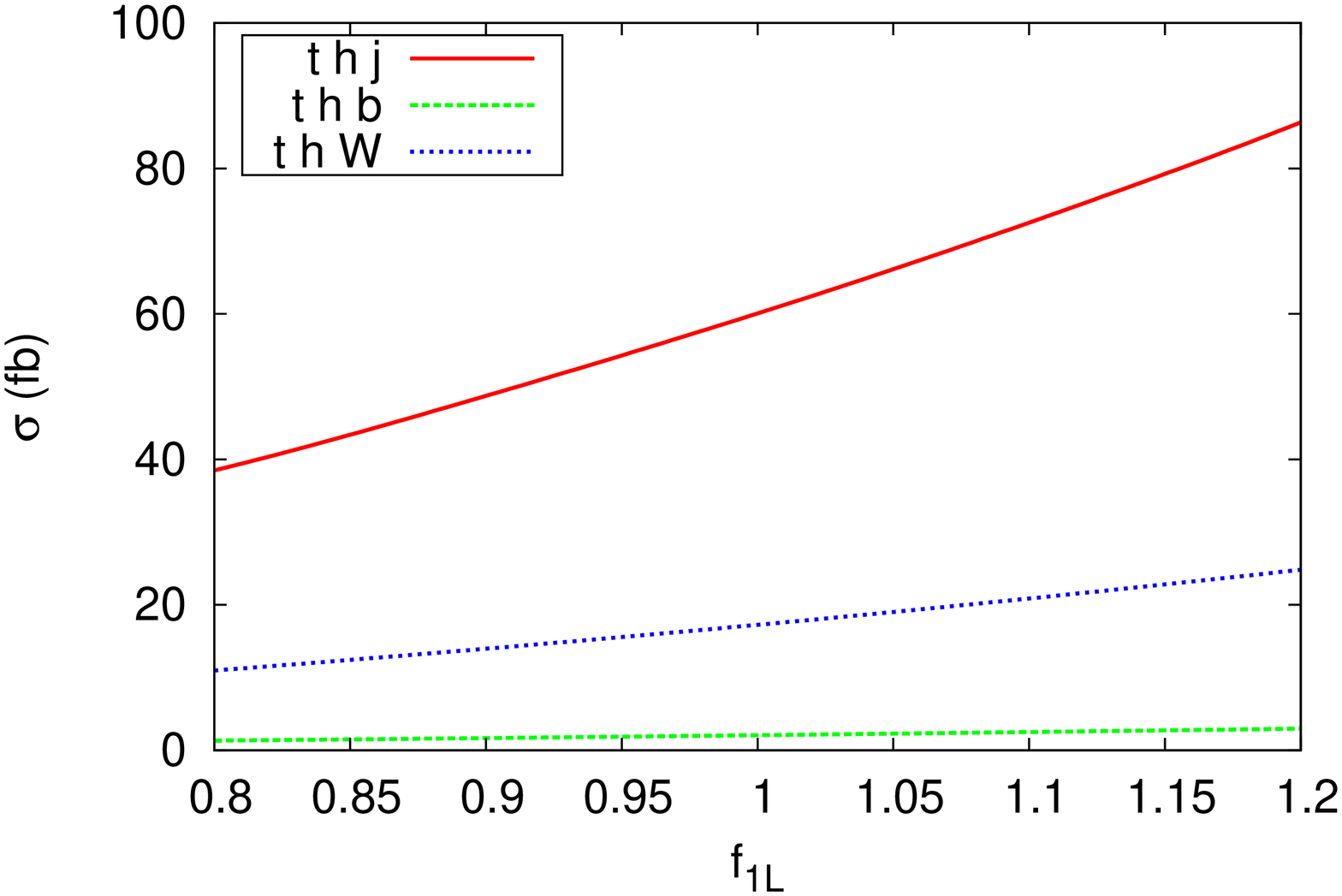}\label{fig:3BFS-f1L}}
\subfigure[]{\includegraphics [angle=-90,width=.49\linewidth] {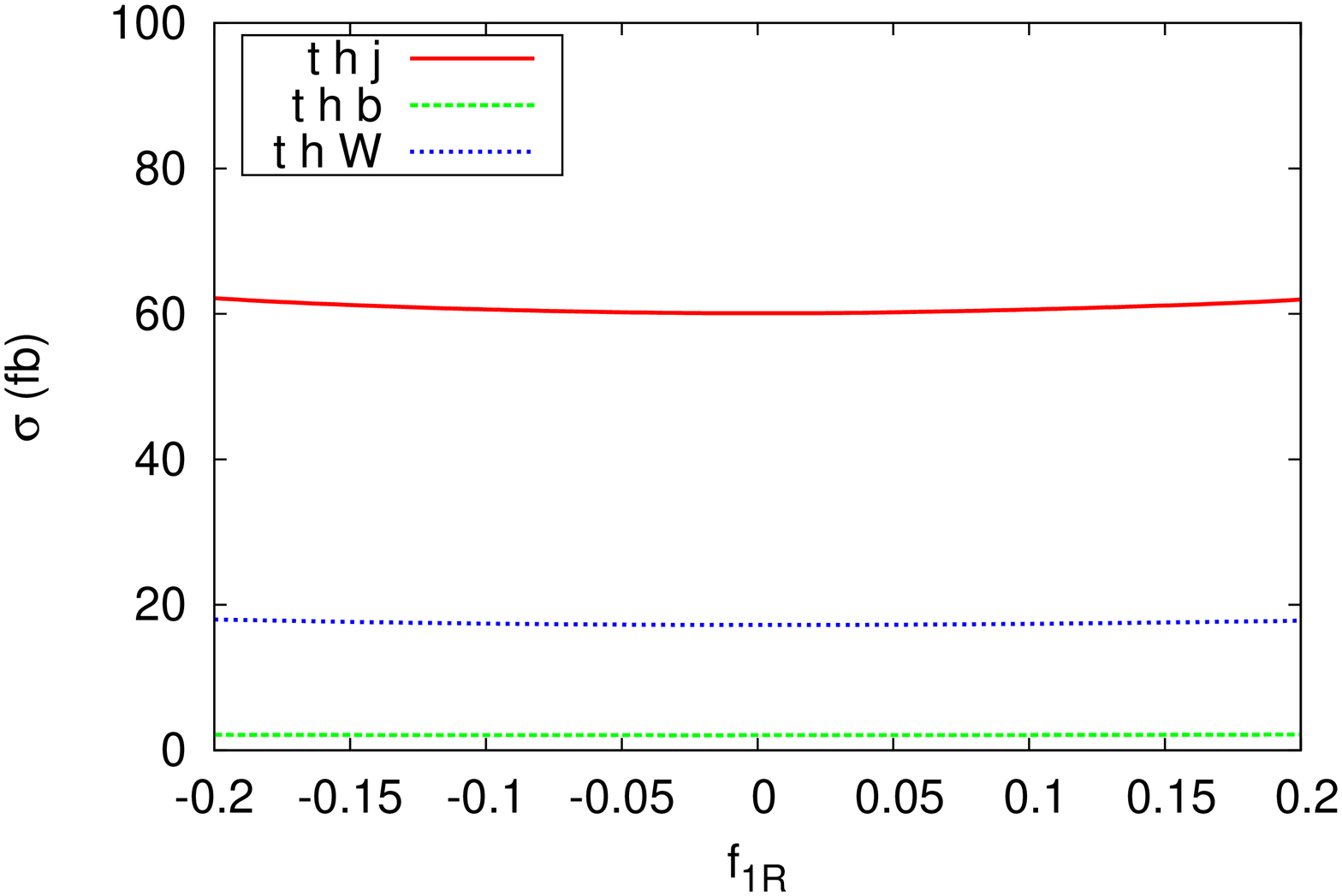}\label{fig:3BFS-f1R}}\\
\subfigure[]{\includegraphics [angle=-90,width=.49\linewidth] {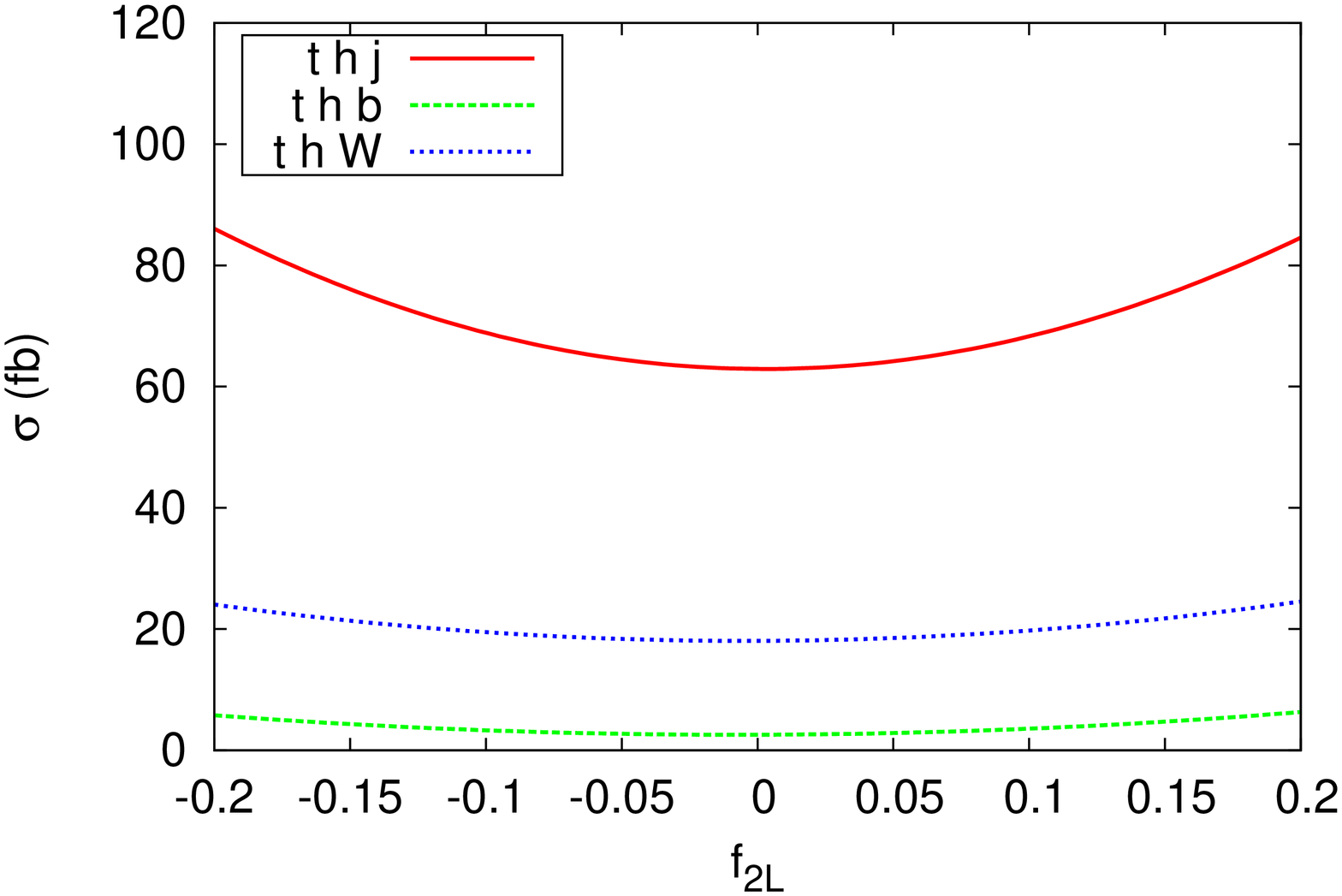}\label{fig:3BFS-f2L}}
\subfigure[]{\includegraphics [angle=-90,width=.49\linewidth] {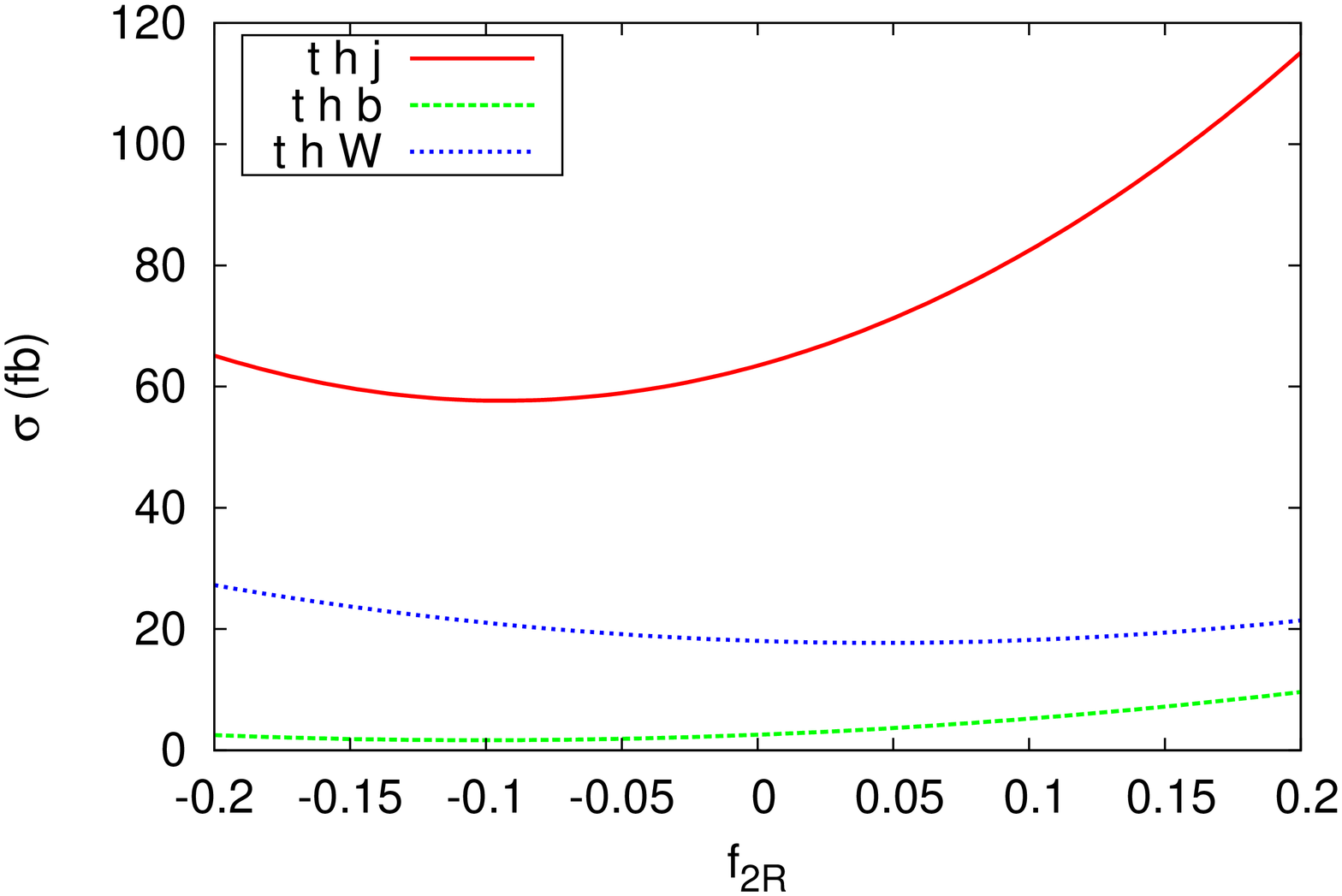}\label{fig:3BFS-f2R}}\\
\subfigure[]{\includegraphics [angle=-90,width=.49\linewidth] {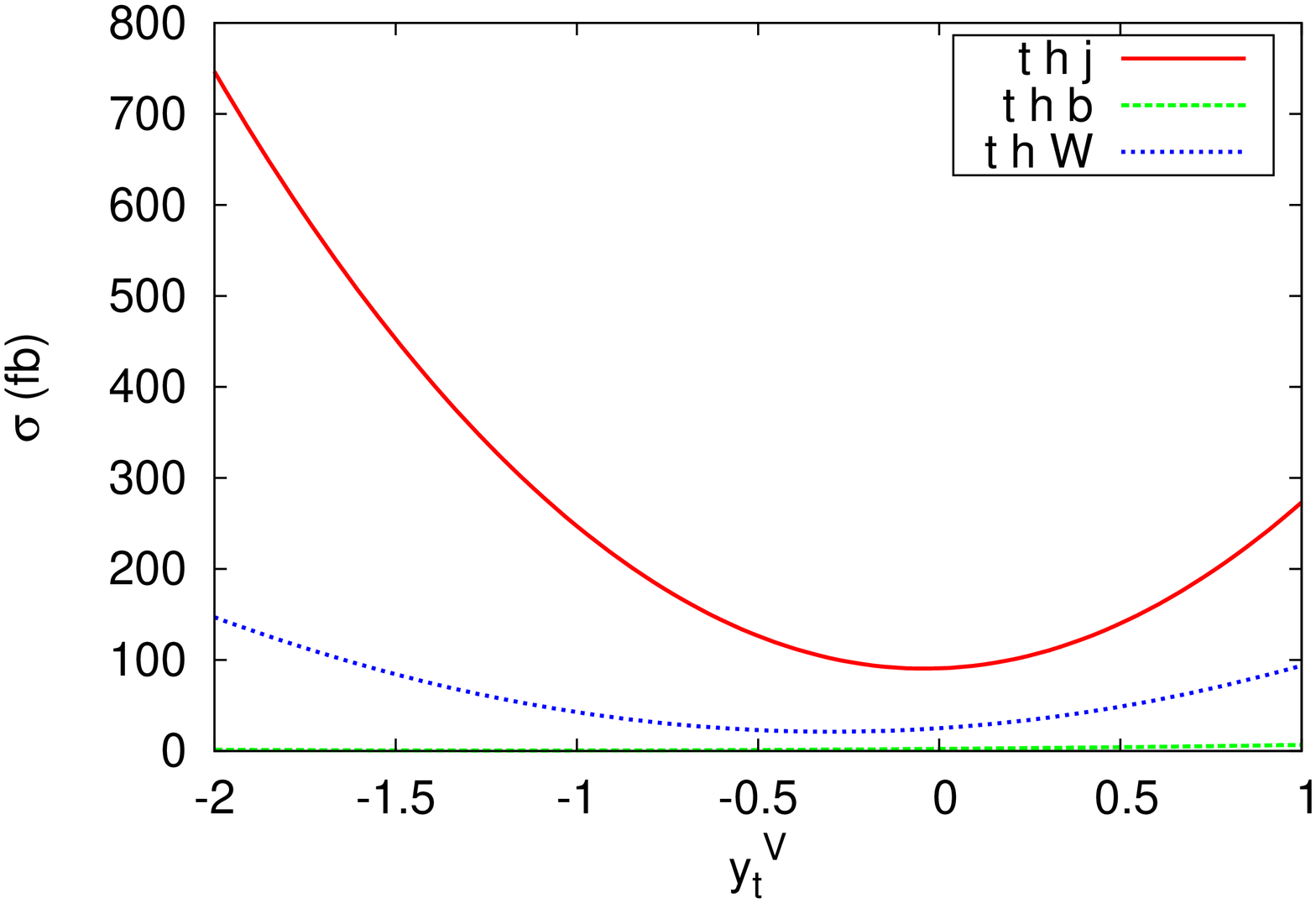}\label{fig:3BFS-ytR}}
\subfigure[]{\includegraphics [angle=-90,width=.49\linewidth] {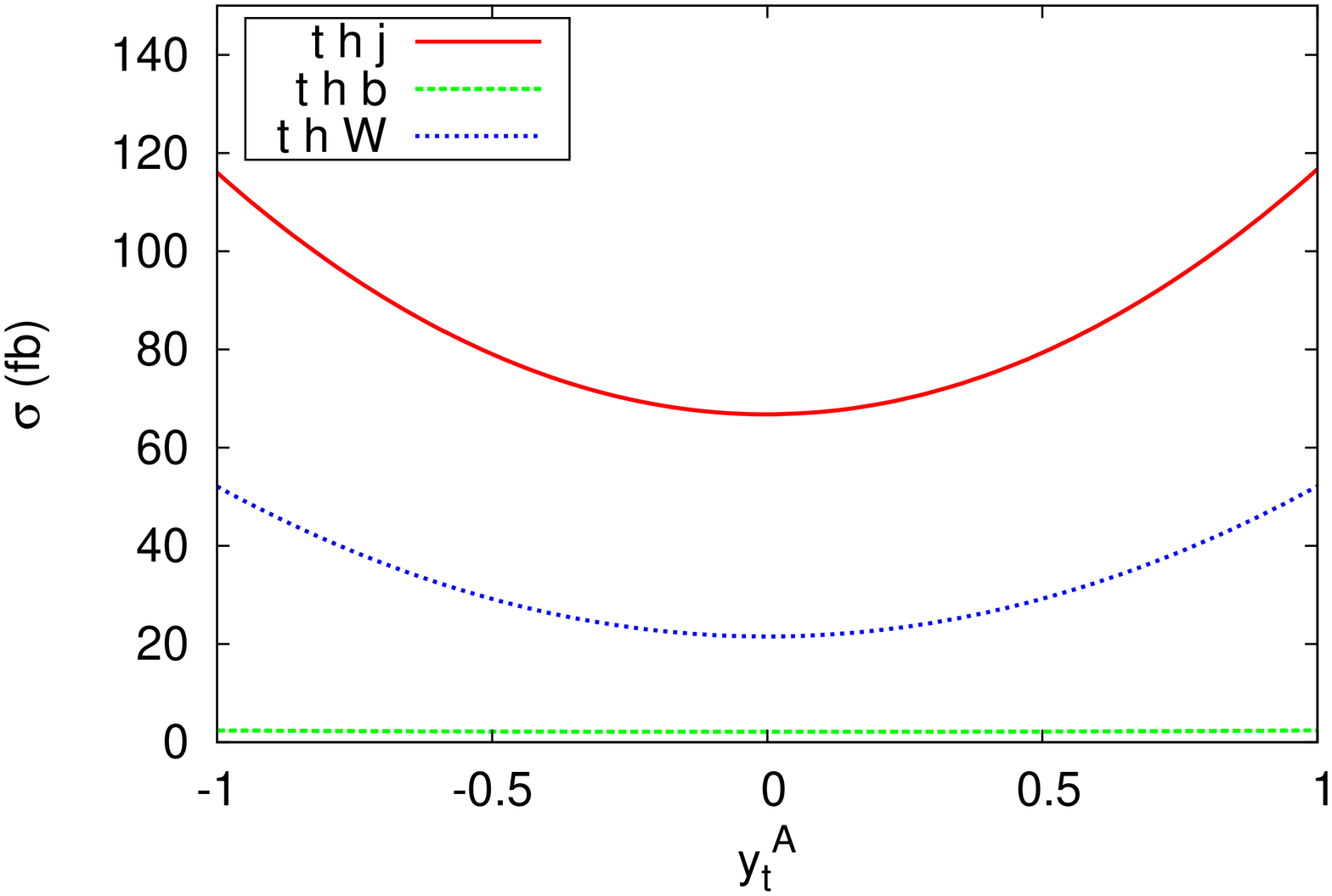}\label{fig:3BFS-ytI}}
\ec
\caption{Dependence of the leading order partonic cross section on $f_{1L}$, $f_{1R}$,$f_{2L}$, $f_{2R}$, $y_t^V$, $y_t^A$. 
Here the individual contribution of the three separate subprocesses are marked by the final state particles. Eq. \ref{eq:cuts}
shows the cuts used on the final state partons. }
\label{fig:3BFS}
\end{figure}

\begin{figure}[!h]
\bc
\subfigure[]{\includegraphics [angle=-90,width=.49\linewidth] {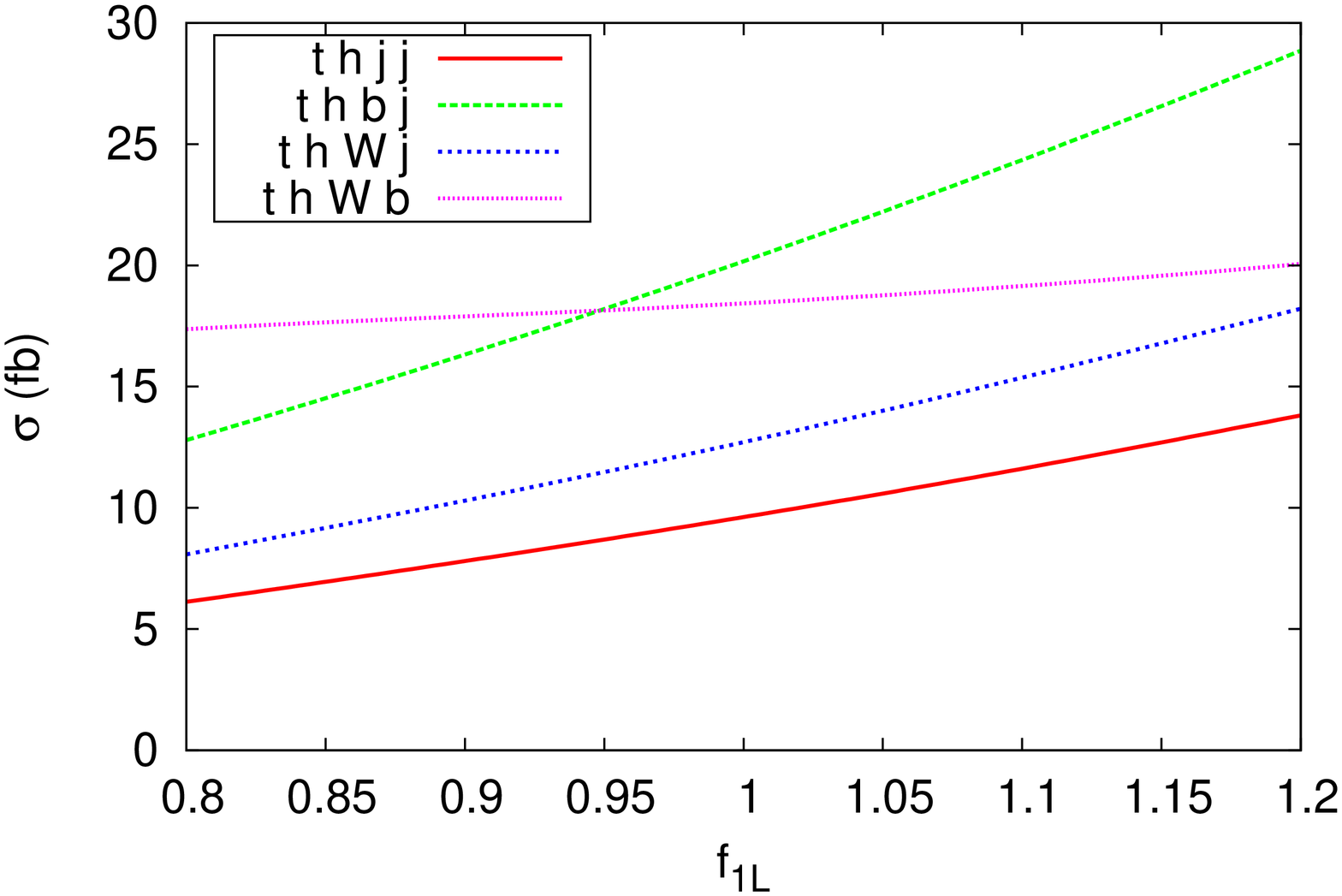}\label{fig:4BFS-f1L}}
\subfigure[]{\includegraphics [angle=-90,width=.49\linewidth] {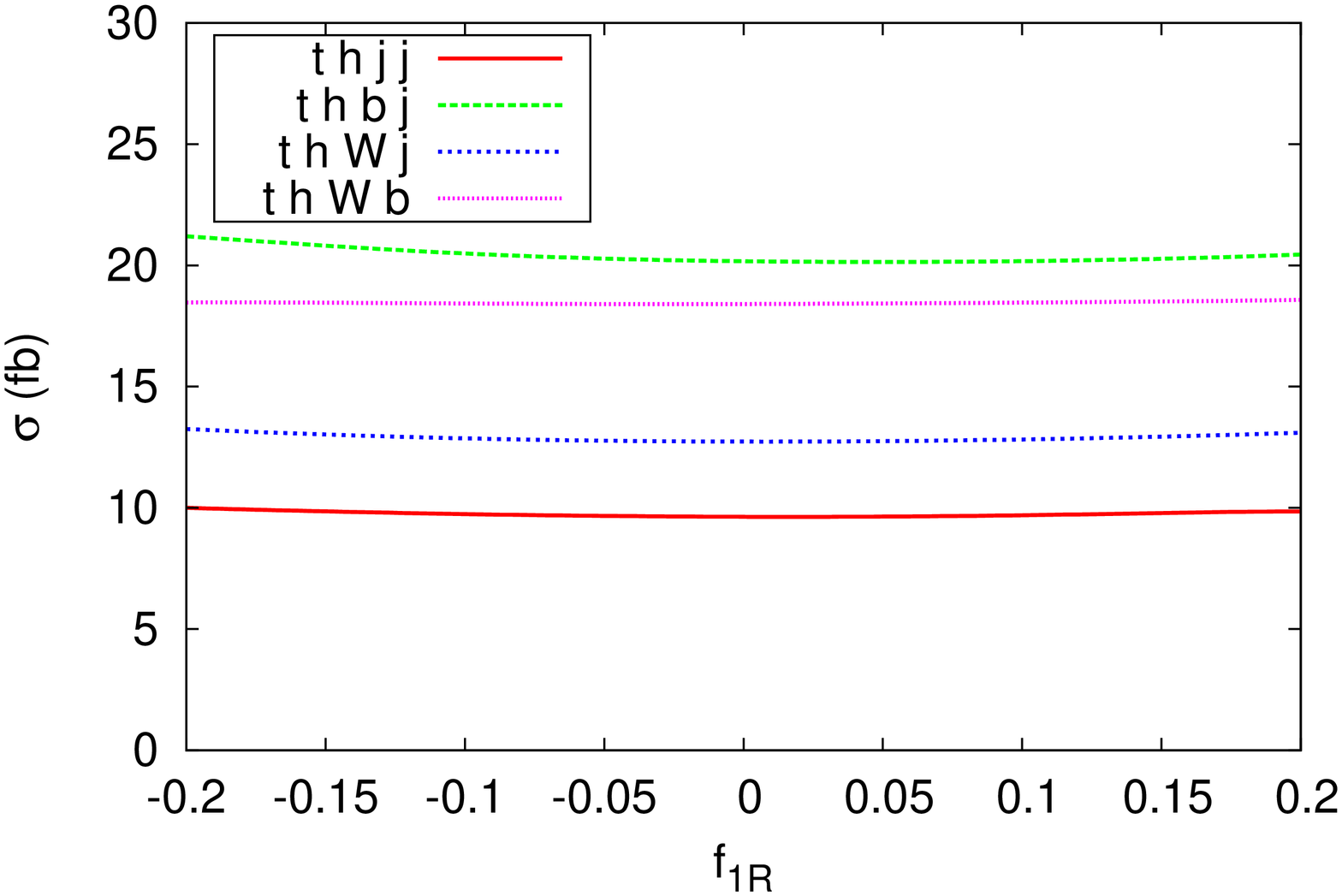}\label{fig:4BFS-f1R}}\\
\subfigure[]{\includegraphics [angle=-90,width=.49\linewidth] {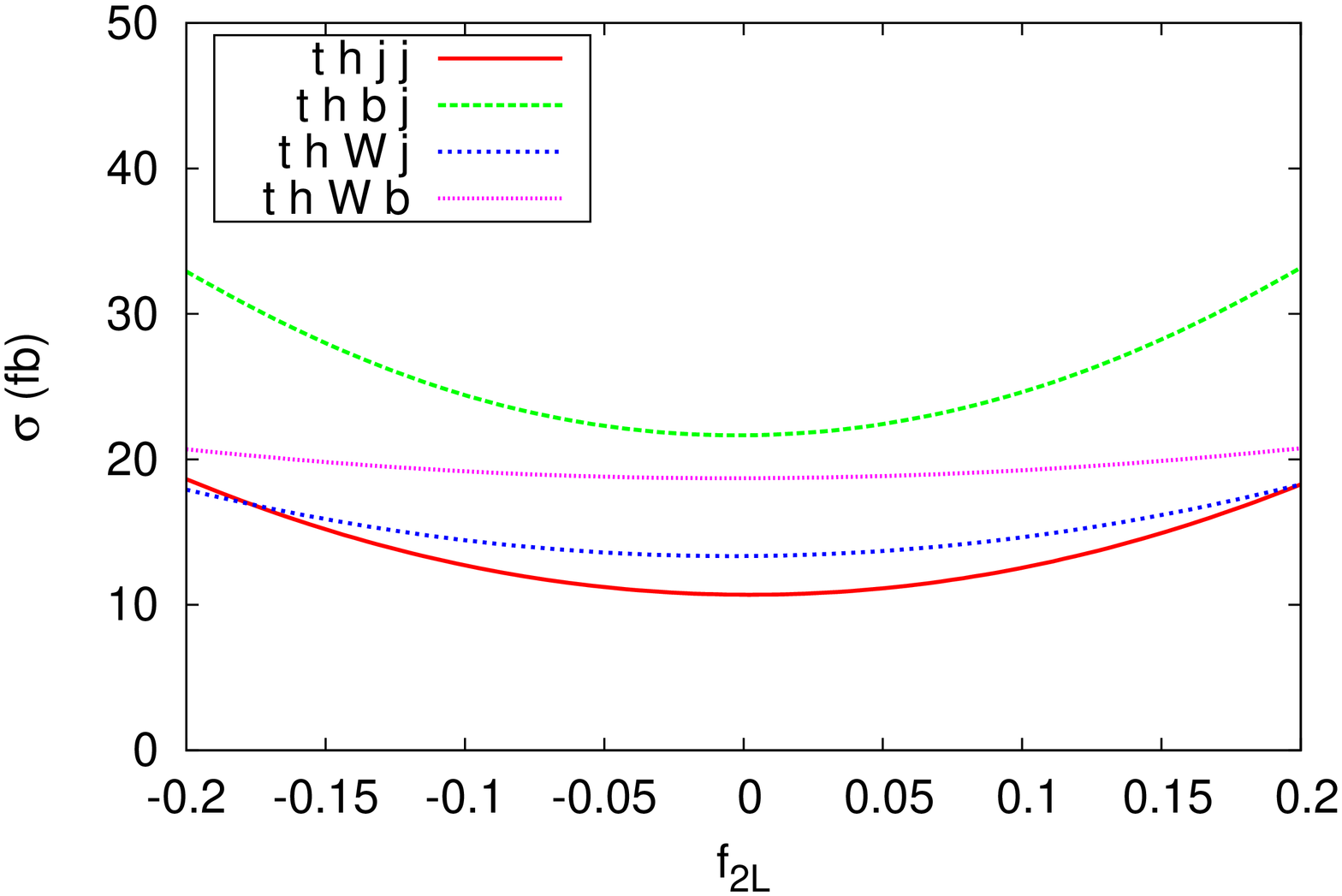}\label{fig:4BFS-f2L}}
\subfigure[]{\includegraphics [angle=-90,width=.49\linewidth] {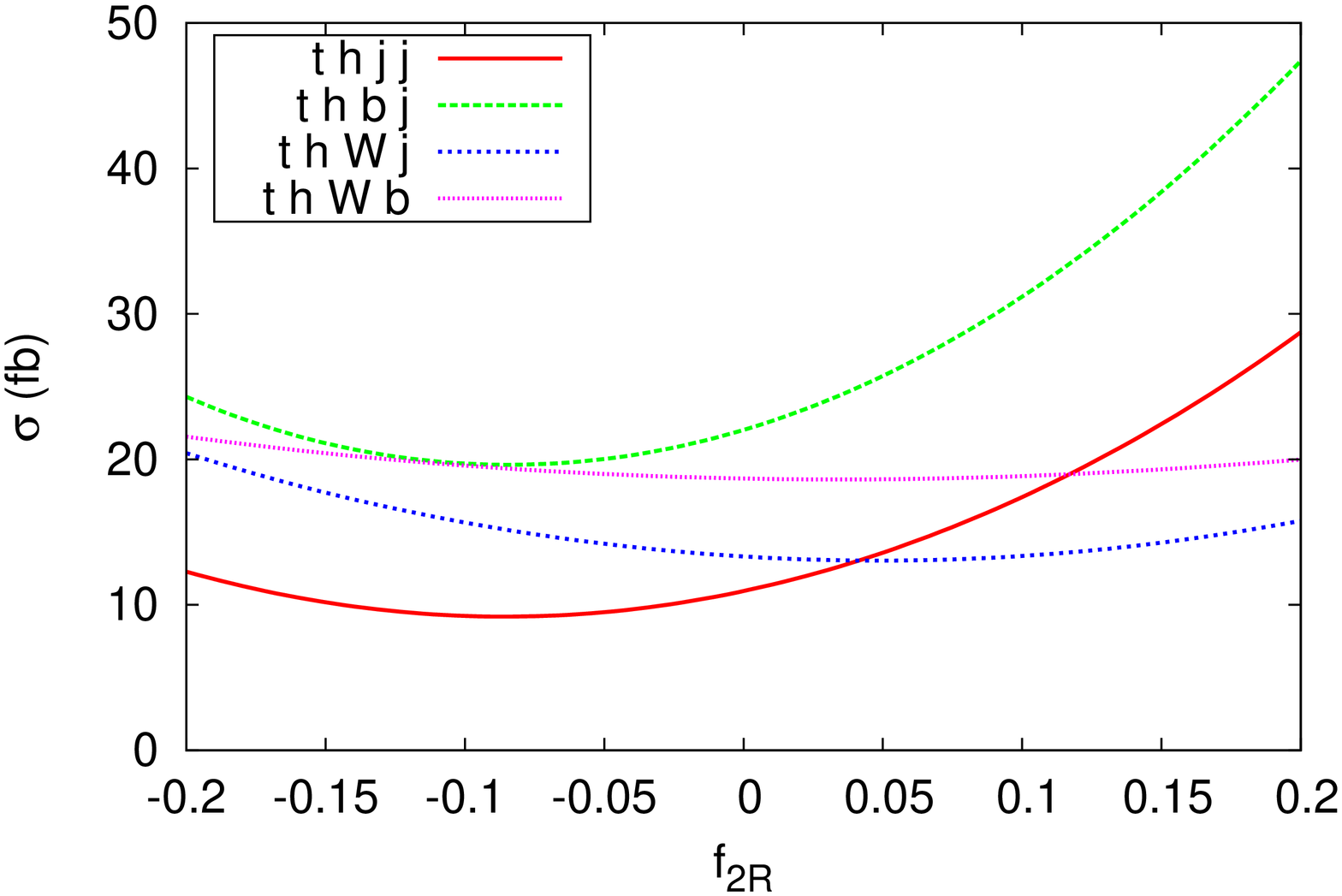}\label{fig:4BFS-f2R}}\\
\subfigure[]{\includegraphics [angle=-90,width=.49\linewidth] {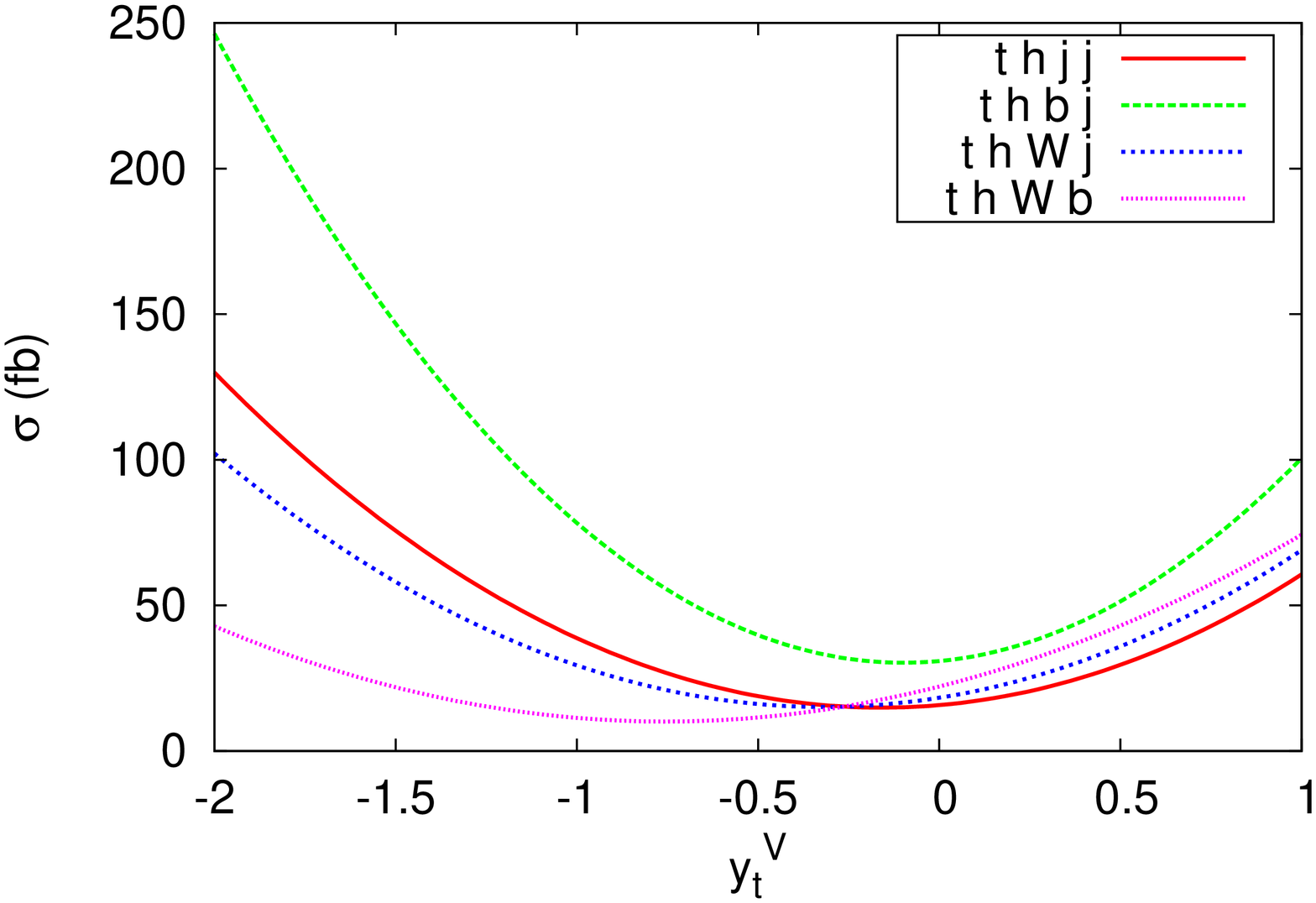}\label{fig:4BFS-ytR}}
\subfigure[]{\includegraphics [angle=-90,width=.49\linewidth] {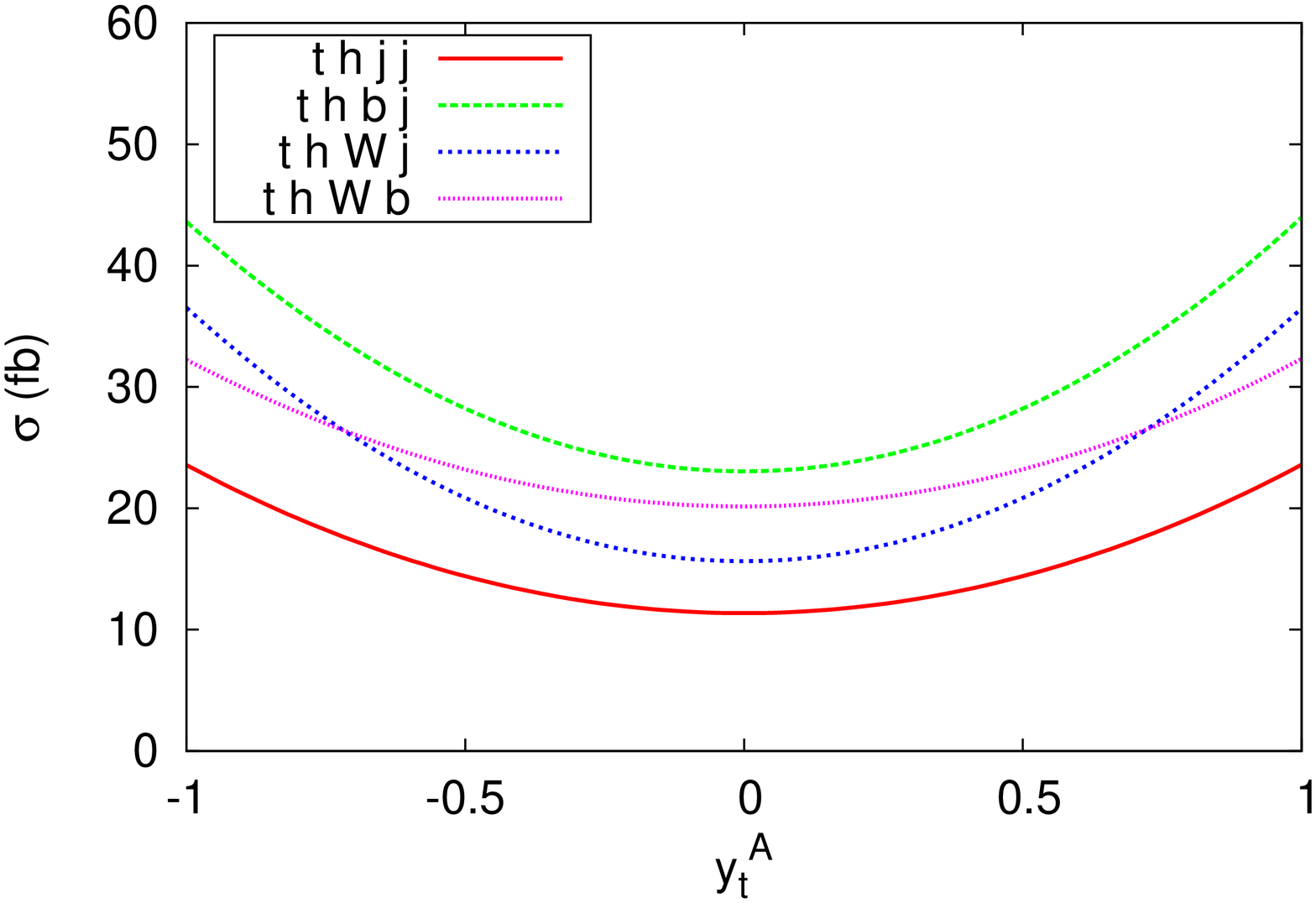}\label{fig:4BFS-ytI}}
\ec
\caption{Dependence of the partonic cross section for processes with 4 particles in the final states on 
$f_{1L}$, $f_{1R}$,$f_{2L}$, $f_{2R}$, $y_t^V$, $y_t^A$. 
The individual contribution of the separate subprocesses are marked by the final state particles. 
Here $j$ stands for a light jet. Eq. \ref{eq:cuts}
shows the cuts used on the final state partons. }
\label{fig:4BFS}
\end{figure}

To compute the cross sections for the processes involved, we first implement the new couplings in \textsc{FeynRules} \cite{Christensen:2008py} and then use \textsc{Madgraph5} \cite{Alwall:2011uj} 
with LO CTEQ6L1 parton distribution functions \cite{Pumplin:2002vw}. We have used the 
following set of kinematic cuts on the final state partons,
\ba
p_T^J > 30 \; \rm{GeV},\: |\eta_J| < 5.0,\; \Dl R(J_1,J_2) = \sqrt{\left(\Dl \et_{J_1,J_2}\right)^2 +\left(\Dl \ph_{J_1,J_2}\right)^2} > 0.4 \label{eq:cuts}
\ea
where $J$ denotes either a light jet or a $b$-jet.
Unlike the $tbW$ and $tth$ anomalous couplings, we find that the associated production of a
single top quark with a Higgs boson is less sensitive to any variation of $WWh$ anomalous 
couplings. If one varies $g^i_{Wh} (i=1,2)$ within the ranges shown in Eqs. \ref{eq:lmtwwh1} 
and \ref{eq:lmtwwh2}, the cross sections for the different processes vary marginally,
about 10-20\%; the production of $thb$ is an exception that can increase by about 60\%. The variation 
of $g^3_{Wh}$ has very little impact on the cross sections. 

In Fig. \ref{fig:3BFS}, we show the dependence of the cross sections of the processes $thj,thb$ and $thW$
on $f_{1L}$, $f_{1R}$, $f_{2L}$, $f_{2R}$, $y_t^V$, and $y_t^A$. The SM value of the
cross section for the $thj$ process is about 60 fb. The variation in $f_{1L}$ and $f_{2L}$ does not 
increase the cross section much. However, at the edge of allowed values of $f_{2R}$ cross section can
double. There is almost no change in the cross section on varying  $f_{1R}$. This overall behavior is almost
like that of the top quark width. So, it can be understood similarly. However, 
there is a strong dependence on the Yukawa couplings. As we shall see below, there exist allowed regions
in the phase space where cross section can increase more than 10 times and approaches 600-800 fb. The
cross sections of the other two processes $thb$ and $thW$ do not depend significantly on the
anomalous $tbW$ coupling. However, the cross section of the $thb$ can almost double with the
allowed range of the Yukawa couplings. In Figs. \ref{fig:3BFS-ytR} and \ref{fig:3BFS-ytI}, we can see
the destructive interference between the $WWh$ and $tth$ couplings in the $thj$ production process \cite{Maltoni}.

In Fig. \ref{fig:4BFS}, we show the dependence of the cross sections of the processes $thjj,thbj,thWb$ and $thWj$
on $f_{1L}$, $f_{1R}$, $f_{2L}$, $f_{2R}$, $y_t^V$, and $y_t^A$. The behavior of the $thjj$ and $thbj$ processes
is similar to what we find above. The variation in $f_{1L}$ and $f_{2L}$ changes cross sections marginally;
the variation in  $f_{1R}$ has almost no impact on the cross sections. However, at the edge of the allowed parameter
values of $f_{2R}$, the cross sections can double. The cross sections of the processes $thWb$ and $thWj$ 
have very weak dependence on the anomalous $tbW$ coupling parameters.
However, as earlier, the cross sections have strong dependence on the Yukawa couplings. 

The plots in  Fig. \ref{fig:3BFS} and  Fig. \ref{fig:4BFS} show variation with respect to
change in one parameter, while the other parameters are kept at the SM value. Of course,
we can choose values of all parameters away from the SM values which will give larger 
cross sections. We have chosen a set of values which may favor the larger cross sections.
This set of values and the cross sections for those values are given in Table \ref{tab:extremepars}.
(Some recent analyses indicate that the data actually disfavors some of these parameter 
points \cite{Nishiwaki:2013cma}. We display these points 
in the table for illustration only.) The set of
parameters $\mc P_0$ corresponds to the SM values. The cross sections of the processes are
adding up to about 150 fb. However, there exist parameter sets where the cross sections
can add up to more than $1$ pb. For most of the listed processes, the cross sections can 
increase as much as fifteen times or more. With these values of the cross sections, it may
be possible to isolate the production of the Higgs boson in association with a top
quark from the background and observe it at the LHC. We note that anomalous couplings
will also change the angular distributions of the jet and the Higgs boson. In particular,
we find that anomalous $tbW$ coupling enhances the cross section more in the central-rapidity
region of the jet and the Higgs boson for the $thj$ production.

\begin{table}[!h]
\bc
\begin{tabular}{|c|rrrrrrr|}\hline
Param.&	$\s_{pp\to thj}$ &$\s_{pp\to thb}$&$\s_{pp\to thW}$&$\s_{pp\to thjj}$&$\s_{pp\to thbj}$&$\s_{pp\to thWj}$&$\s_{pp\to thWb}$\\
Set&(fb)&(fb)&(fb)&(fb)&(fb)&(fb)&(fb)\\\hline\hline
$\mc P_0$&  	 59.6&	2.1&	17.1&	9.6&	20.1&	12.7&	18.4\\
$\mc P_1$&65.1	&	2.5&	16.9&	10.7	&	22.4	&	12.4	&	18.4\\
$\mc P_2$&69.2 &	3.5&	19.3&	13.1&		24.2&		14.0&		19.1\\
$\mc P_3$&57.3&2.0&17.1&		9.5&		19.9&		12.7&		18.4\\
$\mc P_4$&	180.1&	2.7&	51.6&	35.1&	72.4&	35.8&	18.3\\
$\mc P_5$&	382.9&	3.2&	105.4&	69.6&	144.3&	73.0&	30.3\\
$\mc P_6$&	472.0&	3.4&	116.7&	86.7&	153.3&	79.9&	32.9\\
$\mc P_7$&	567.0&	53.0&	129.9&	169.0&	246.1&	95.3&	93.5\\
$\mc P_8$&	602.3&	29.4&	250.7&	163.8&	263.3&	184.2&	117.1\\
$\mc P_9$&	875.2&	64.4&	229.8&	241.5&	363.5&	167.0&	107.4\\
\hline
\end{tabular}

\vskip 0.2in

\begin{tabular}{|c|rrrrrrrrr|}\hline
Param. Set&$f_{1L}$ &$f_{1R}$&$f_{2L}$&$f_{2R}$&$y_{t}^V$&$y_t^A$
&$g_{Wh}^1$&$g_{Wh}^2$&$g_{Wh}^3$\\
\hline\hline
$\mc P_0$&  	1.0&0.0&0.0&0.0&0.0&0.0&0.0&0.0&0.0\\
$\mc P_1$&	 	1.0&0.0&0.0&0.0&0.0&0.0&0.1&0.0&0.0\\
$\mc P_2$&	 	1.0&0.0&0.0&0.0&0.0&0.0&0.0&0.3&0.0\\
$\mc P_3$&	 	1.0&0.0&0.0&0.0&0.0&0.0&0.0&0.0&0.2\\
$\mc P_4$&	0.8&0.2&0.2&0.2&-1.0&-1.0&0.0&0.0&0.0\\
$\mc P_5$&	1.2&0.2&0.2&0.2&-1.0&-1.0&0.0&0.0&0.0\\
$\mc P_6$&	1.2&0.2&0.2&-0.2&-1.0&-1.0&0.0&0.0&0.0\\
$\mc P_7$&	0.8&0.2&0.2&0.2&1.0&1.0&0.0&0.0&0.0\\
$\mc P_8$&	1.2&0.2&0.2&-0.2&1.0&1.0&0.0&0.0&0.0\\
$\mc P_9$&	1.2&0.2&0.2&0.2&1.0&1.0&0.0&0.0&0.0\\\hline
\end{tabular}
\caption{\label{tab:extremepars}Cross-sections for different single top quark and Higgs boson associated production processes for six different 
choices of anomalous coupling parameters denoted by $\mc P_{i=1,\ldots,9}$ (explained in the lower table). The set $\mc P_0$ corresponds to
the SM couplings while in sets $\mc P_{1,2,3}$ only $g_{Wh}^{1,2,3}$s are varied.}
\ec
\end{table}

\section{Observability}

   We now consider the possible signatures of these processes and their dominant 
   backgrounds to show that the backgrounds to some of the processes can be manageable.

 For $m_h \approx 125$ GeV, the primary decay mode of the Higgs boson 
    is $h \to b {\bar b}$.  To observe any signature of the processes, the 
    accompanying top quark needs to decay semi-leptonically.
    If it decays into jets, the QCD backgrounds from various multijet events would overwhelm the signal. 
    A very simple signature for all the processes would be ``an isolated $e/\mu$ + jets'', where the top quark decays semi-leptonically and the other particles are either jets or decay into
    jets. Such a signature would not be viable due to very large background
    from the processes such as ``$W$ + jets'' and ``$t$ + jets''. 
    However, since most of the jets in the signal
    processes are $b$-jets, we can use the tagging of the $b$-jets
    to reduce the backgrounds. In particular, for the signature --
    ``an isolated $e/\mu$ + 3 $b$-jets + light jets'' \cite{Farina:2012xp}, all of the processes under consideration can contribute. To isolate different signal processes, one has to look for other signatures. For example, a signature specific to $tbh$ and $tbhj$ is ``isolated $e/\mu$ + 4 $b$-jets''.
Similarly, ``2 isolated $e/\mu$ + 3 $b$-jets'' can come from the $W$ boson associated 
productions {\it i.e.}, $thW$, $thWj$ and $thWb$ when the $W$ boson also decays into leptons.  
Since there is an extra $b$-quark in the $thWb$ production, one can also consider 
 ``2 isolated $e/\mu$ + 4 $b$-jets'' to isolate this process.  In this paper, we  investigate some of these signatures and the corresponding backgrounds in detail and  estimate the statistical significance of the signal over background for each of these signatures. 
For the signal we consider three cases with three different sets of anomalous couplings consistent with the currently available bounds.
    \begin{itemize}
    \item Case 1: we consider maximally allowed anomalous $tth$ coupling only \cite{Nishiwaki:2013cma} -
    $f_{1L} = 1.0, f_{1R} = f_{2L} = f_{2R} = 0, y_{t}^V = -1.5, y_t^A = 0.5$.
    \item
    Case 2: we consider almost maximally allowed anomalous $tbW$ coupling only -
    $f_{1L} = 1.2, f_{1R} = f_{2L} = 0, f_{2R} = 0.2, y_{t}^V = 0, y_t^A = 0$.
    \item
     Case 3: we consider the combination of the above two cases -
    $f_{1L} = 1.2, f_{1R} = f_{2L} = 0, f_{2R} = 0.2, y_{t}^V = -1.5, y_t^A = 0.5$.
    \end{itemize}

Like the signal, we generate events for the potentially significant background processes
(both irreducible and reducible) at the parton level with \textsc{MadGraph5}.  When a background process also includes $tbW$ and/or $tth$ vertices, we compute it separately for the three cases mentioned above. Since, this is a
    parton level study, it is important to include appropriate 
    smearing of the parton energies to simulate the energy resolution of a jet. 
    We use the following resolution function,
    $$
     {\Delta E \over E} = {a \over E} + {b \over \sqrt{E}} + c.
    $$
      For a parton jet, we take $a = 4.0, b = 0.5, c = 0.03$. We also smear
    the energy of an electron/muon with $a = 0.25, b = 0.1, c = 0.007$. Here, $E$ is in the
    units of GeV.  We then
    construct the smeared four-momenta of the particles using this smeared energy.
 We have taken the efficiency
    of identifying a $b$-jet as $60\%$. For the reducible backgrounds, we consider the possibility of a light jet to be mistagged as a $b$-jet. For this, the mistagging efficiency for a charm quark is taken as $10\%$ and for any other quark/gluon it is $1\%$. 
The choice of smearing parameters and the tagging and mistagging efficiencies
    are more or less consistent with the ATLAS experiment.

    In Table \ref{tab:l3bj}, we display the results for ``an isolated $e/\mu$
    + 3 $b$-jets + a light jet'' which is a signature for the $pp \to thj$ process. 
    For the backgrounds (here and below), we consider only the significant ones. 
    For all the cases, we apply the following generic cuts:
\ba
    p_T^{b,\ell} > 20 \; {\rm GeV},\: |\eta_{b, \ell}| < 2.5,\;    p_T^j > 25 \; {\rm GeV},\: |\eta_j| < 4.5,\; \Dl R(J/\ell,J/\ell) > 0.4 \label{eq:cuts1}.
\ea
     In addition, we require $|M(bb) - M_h| < 15\;$ GeV for at least one $b$-jets pair. 
     In Cases 1 and 3, we also require the
     light jet to be forward, i.e., $ |\eta_j| > 2.5$. There is also a requirement for the minimum
     $M(jb)$ for all pairs. Its value for Cases 1, 2 and 3 are 100 GeV, 50 GeV and 90 GeV
     respectively. Specially for Case 2, where the background is relatively larger and signal smaller compared to the other two cases, we also require $M(jbb) > 220$
     GeV for all combinations and  $M(ljb) > 290$ GeV for only highest $p_T$ $b$-jet.

\begin{table}[!h]
\begin{center}
\begin{tabular}{|c|r r|rrrrrrr|rr|}  \hline
& \multicolumn{2}{c|}{Signal} & \multicolumn{7}{c|}{Backgrounds} & \multicolumn{2}{c|}{$S/\sqrt{B}$}  \\ \cline{2-12}
        &SM    & Ano. &$tZj$  &$tbbj$  &$Wbbbj$ &$tt$   &$ttj$    &$tbjj$ &$Wbbjj$  &SM   &Ano.  \\  \hline\hline
Case 1  &46.45 & 536.68    &23.59  &65.39   &11.10  &0.00   &6129.60  &191.81 &92.74    &0.58 &6.65   \\
Case 2  &74.04 & 187.98    &158.87 &139.27  &42.07  &0.00   &16524.10 &748.22 &262.90   &0.55 &1.41   \\
Case 3  &48.91 & 702.35    &107.51 &106.18  &12.28  &15.01  &6436.08  &340.34 &99.89    &0.58 &8.33    \\ \hline
\end{tabular}
\end{center}
\caption{\label{tab:l3bj}Number of events for the signature ``an isolated $e/\mu$
    + 3 $b$-jets + a light jet''
at the 14 TeV LHC with the integrated luminosity of 100 fb$^{-1}$. The cuts and efficiencies are specified in the text.}
\end{table}     
     
     We see that with 100 fb$^{-1}$ integrated luminosity, the signal significance for the pure SM is too low to be observed with such a simple kinematical cut based analysis. \footnote{For low statistics, especially when $S > B$, the ratio $S/\sqrt{B}$ overestimates the signal significance. In that case, one may switch to the quantity $\sqrt{2(S+B){\rm ln}(1+S/B) -2S}$ for significance estimation \cite{Cowan:2010js}. } Here, a multivariate analysis may improve the statistics. Also, for Case 2, after the specialized cuts the signal significance is still not as good as the other two. 
The results indicate that with this signature, even with the maximally allowed anomalous $tbW$ couplings the signal can only be detected after the end of the second LHC run if the integrated luminosity is large enough, but, one can put some bounds within a year of the LHC restart on the anomalous $tth$ couplings. However, as we will see below, there are better signatures to probe these couplings.


    In Table \ref{tab:l4bj}, we display the results for ``isolated $e/\m$ + 4$b$-jets + a light (forward) jet'' -- a signature for the $thbj$ signal. If we don't include the light jet in the signature, the signal will also get contribution from the $thb$ process. However, for the values of the anomalous couplings that we consider, the process $thb$ has
    very small cross section, even with the maximal anomalous couplings. Therefore, we don't include its contribution and include the forward light jet in the signature which can help to reduce the background. For all the cases, we apply the same generic cuts as in  Table 2.
     In addition, we require $|M(bb) - M_h| < 15\;$ GeV, $ |\eta_j| > 2.0$, $M(bb) > 100$ GeV for all
    pairs of $b$-jets, and $M(bj) > 150$ GeV for all pairs. Specifically for Case 2, we also apply a cut on $M(bj)$ on all $bj$ pairs except for the smallest $p_T$ $b$-jet and $M(bb) > 120$ GeV.

\begin{table}[!h]
\begin{center}
\begin{tabular}{|c|rr|rrrrrr|rr|} \hline
& \multicolumn{2}{c|}{Signal} & \multicolumn{6}{c|}{Backgrounds} & \multicolumn{2}{c|}{$S/\sqrt{B}$}  \\ \cline{2-11}
         &  SM & Ano. & $tZbj$ & $tbbbj$ & $ttbb$ & $tth$ & $ttj$ &$tbbjj$ & SM & Ano. \\ \hline\hline
Case 1   &3.26 & 33.53     & 0.21   & 2.32   & 0.23   & 0.03  & 0.03  & 0.07   &1.92  & 19.72 \\
Case 2   &2.60 & 6.86      & 0.69   & 2.41   & 0.71   & 0.46  & 0.00  & 0.02   &1.26  &  3.31   \\
Case 3   &3.26 & 49.52     & 3.41   & 4.88   & 0.00   & 0.08  & 0.03  & 0.05   &1.12  & 17.03 \\ \hline
\end{tabular}
\end{center}
\caption{\label{tab:l4bj}Number of events for the signature ``isolated $e/\m$ + 4$b$-jets + a light (forward) jet''
at the 14 TeV LHC with the integrated luminosity of 100 fb$^{-1}$.
The cuts and efficiencies are specified in the text. 
The reducible background $tbbjj$ includes $ttb$.}
\end{table}

     Our choice for the cuts is not necessarily optimum. Rather, it is to illustrate that 
     anomalous couplings can show up in the associated production of the single top
     quark and a Higgs boson. We see that to probe the anomalous couplings this signature is better than the earlier one as the signal significances are better in all the three cases. Because of the larger enhancement of the
     cross sections due to the anomalous $tth$ couplings, the signal for the maximal couplings 
     would be visible within a few months of the restart of the LHC. Even much smaller enhancement
     of the cross section, say lower by a factor of 5-6 would also show up in the
     second run of the LHC. It will, however, take more than a year to see the signal if only $tbW$ 
     couplings are anomalous. One can also look for other strategies to enhance the significance
     in this case.
     For example, we find that if we drop the requirement of the light jet being
     a forward jet and require a minimum $M(bj)$ for all pairs, then it is possible
     to increase the significance to almost 4.

    In Table \ref{tab:2l3bj}, we display the results for the signature ``2 isolated $e/\mu$ + 3 $b$-jets'' for $thW$ process.
    If we allow an extra light jet in the signature then both $thW$ and $thWj$ will contribute to the signal.
Here, however, for simplicity, we don't demand the extra light jet in the signature and display the results for the $thW$ signal process only.
    Like before, we apply the following generic cuts:
    \ba
    p_T^{b,\ell} > 20 \; {\rm GeV},\: |\eta_{b, \ell}| < 2.5,\; \Dl R(J/\ell,J/\ell) > 0.4 \label{eq:cuts2}.
\ea
     In addition, we require $|M(bb) - M_h| < 15\;$ GeV, $M(\ell b) > 180$ GeV for all
     pairs of a lepton and a $b$-jet. Since we are now demanding 2 leptons in the final state, a potentially large background can come from ``$Z/\g^*$ + jets" processes.
However, the requirement of three $b$-tagged jets and the invariant mass
cuts described above makes this background small. Moreover, it is possible to almost eliminate the ``$Z$ + jets" background with suitable cuts on the invariant mass
of the lepton pair. Hence, we don't include this background in our estimation.
     
\begin{table}[!h]
\begin{center}
\begin{tabular}{|c|rr|rr|rr|} \hline
& \multicolumn{2}{c|}{Signal} & \multicolumn{2}{c|}{Backgrounds} & \multicolumn{2}{c|}{$S/\sqrt{B}$}  \\ \cline{2-7}
         & SM & Ano. & $ttb$ & $ttj$ & SM & Ano. \\ \hline\hline
Case 1   &  0.65 & 8.01 & 0.09 &  0.14 & 1.36 & 16.40 \\
Case 2   &  0.65 & 1.06 & 0.00 &  0.14 & 1.74 & 2.80   \\
Case 3   &  0.65 & 11.60 & 0.00 &  0.14 & 1.74 & 30.58 \\ \hline
\end{tabular}
\end{center}
\caption{\label{tab:2l3bj}Number of events for the signature ``2 isolated $e/\mu$ + 3 $b$-jets''
at the 14 TeV LHC with the integrated luminosity of 100 fb$^{-1}$. The cuts and efficiencies are specified in the text.}
\end{table}

      We again clearly see that if $tth$ coupling is anomalous, then within a few months,
      and if $tbW$ coupling is anomalous, then in 2-3 years, the single top quark production
      with a Higgs and a $W$ boson would be visible. Alternatively, one can put quite strong bounds on the anomalous couplings (especially the $tth$), if the signal is not visible.

      Finally, to complete our analysis, we display the results for the signature ``2 isolated $e/\m$ + 4$b$-jets'' for the $thWb$ process 
      in Table \ref{tab:2l4bj}. Event selection cuts are similar to the previous case except for a minimum cut on $M(\ell b)$
      for all the bottom jet and lepton pairs as $M(\ell b)> 160$ GeV in all the cases. We can further reduce the backgrounds 
      without loosing much signal events by making this cut stronger.
      Due to very small cross section of the signal, very large luminosity will be required 
      to observe it at the LHC.
      
\begin{table}[!h]
\begin{center}
\begin{tabular}{|c|rr|rrrrr|rr|} \hline
& \multicolumn{2}{c|}{Signal} & \multicolumn{5}{c|}{Backgrounds} & \multicolumn{2}{c|}{$S/\sqrt{B}$}  \\ \cline{2-10}
         & SM    & Ano. & $ttbb$ & $tth$ & $ttZ$ & $ttbj$ & $ttjj$ & SM   & Ano. \\ \hline \hline
Case 1   &  1.64 &  9.30     & 1.57   & 0.14  & 0.10  & 0.03   & 0.08   & 1.18 & 6.72 \\
Case 2   &  1.64 &  2.90     & 3.74   & 0.72  & 0.11  & 0.06   & 0.13   & 0.75 & 1.33  \\
Case 3   &  1.64 & 13.55     & 3.74   & 0.34  & 0.14  & 0.12   & 0.26   & 0.76 & 6.33  \\ \hline
\end{tabular}
\end{center}
\caption{\label{tab:2l4bj}Number of events for the signature ``2 isolated $e/\m$ + 4$b$-jets''
at the 14 TeV LHC with the integrated luminosity of 1000 fb$^{-1}$. The cuts and efficiencies are specified in the text.}
\end{table}
     
      Other two important decay modes of the Higgs boson, for mass around 
      $125$ GeV, are $h \to \tau \tau, W W^{*}$.  Both have branching ratios 
      of few percents.  Here the decay mode $h \to \tau \tau$ can be useful 
      with the detection of tau-jets. Then a signature of the type 
      ``isolated lepton + 2 tau-jets + $1/2$ bottom jets'' can be useful. The mimic
      backgrounds would be same as that for  $h \to b {\bar b}$ case. Here we will 
      have to include the probability of a jet faking a tau-jet instead of a
      bottom-jet. At a longer time scale even $h \to  W W^{*}$ can also be useful if one looks at 
      ``one/two isolated leptons + two-tau jets + $1/2$ bottom jet''. A more
      detailed study is required for analyzing these signatures.
      
      Before we present our conclusions we would like to note that it may also be possible 
      to obtain good signal significance by considering a signature that is common to all the signals,
      {\it e.g.}, ``$e/\mu$ + 3 $b$-jets + any number of light jets''. As mentioned earlier, in this case all the $pp\to thX$ processes will contribute. However, in this case,
      due to jet multiplicity, a parton level estimation for the backgrounds, such as we do in this paper, may not be appropriate.

\section{Conclusions}

      In this paper, we have investigated the effect of anomalous couplings in the 
      $tbW$, $tth$ and $WWh$ vertices on the associated production of a single top 
      quark with a Higgs boson. We have considered the production of 
      $t h j$, $t h b$, $t h W$, $t h j j$, $t h j b$, $t h W j$, $t h W b$.
      Within the SM, these processes have small cross 
      sections. However, we find that anomalous $Wtb$ and $tth$ couplings can 
      enhance the cross sections of the some of these processes significantly. 
      The cross sections of these processes are mainly sensitive to the top Yukawa 
      couplings and  $f_{1L}$, $f_{2R}$. For some combinations of these couplings, 
      the cross section of some of the processes can be enhanced 
      by more than a factor of 10. The combined cross section of the processes 
      under consideration can be more than $500$ fb. Anomalous $WWh$ couplings 
      plays less significant role; it can mostly enhance the cross sections to 
      the extent of $10 - 20\%$. As a result of the sensitivity to the anomalous top Yukawa couplings and $f_{1L}$, $f_{2R}$, these processes have the 
      potential to act as probes for these couplings.
      
      To verify that these processes can indeed be useful to probe the anomalous couplings at 
      the LHC, we have also done a signal vs. backgrounds
      study with three different choices of the couplings along with the SM case. We have analyzed  the following signatures  -- 
      a)``an isolated $e/\mu$  + 3 $b$-jets + a light jet'' for the $pp\to thj$ process,
      b)``an isolated $e/\mu$  + 4 $b$-jets + a forward light jet'' for the $pp\to thbj$ process,
      c) ``2 isolated $e/\mu$ + 3 $b$-jets" for the $pp\to thW$ process
      and d) ``2 isolated $e/\mu$ + 4 $b$-jets'' for the $pp\to thWb$ process. Our computation  clearly shows that, except the last 
      one,  it is possible to observe these signatures in the next run of the LHC. The last signature suffers from small signal cross section and as a result will require very large luminosity to be observed. In general we find that for large anomalous top Yukawa 
      couplings these signatures will be visible within a year but for purely anomalous $tbW$
      couplings it can take longer unless some other search strategies are used.
   In case the signal is not visible, quite strong bounds
      on the anomalous couplings can be put.  
      
      Finally, we note that if such larger than the 
      SM cross sections are indeed observed in the future, then it would require 
      further analysis to identify the couplings responsible for 
      the enhancement as well as a realistic model that can
      contribute to the enhancement of the cross sections.  
      However, as we saw, there are different viable signatures. So looking at these different signatures together might help in this situation.

\section*{Acknowledgement}
{ AS would like to acknowledge the 
partial support available from the Department of Atomic Energy, Government
of India for the Regional Centre for Accelerator-based Particle Physics (RECAPP), Harish-
Chandra Research Institute.}

\begin{spacing}{1}

\end{spacing}
\clearpage

\end{document}